\documentclass[
  a4paper,
  amsfonts,
  amssymb,
  amsmath,
  reprint,
  showkeys,
  nofootinbib,
  twoside,
  aps,
  pra  % For Physical Review B
]{revtex4-1}

\usepackage[english]{babel}
\usepackage[utf8]{inputenc}  % utf8 is generally preferred over utf8x
\usepackage[T1]{fontenc}

\usepackage{graphicx}
\usepackage{mathtools}
\usepackage[nodayofweek]{datetime}
\usepackage{amsmath}
\usepackage{amssymb}
\usepackage{amsfonts}
\usepackage{amsthm}
\usepackage{url}
\usepackage{graphicx}
\usepackage{dcolumn}% Align table columns on the decimal point
\usepackage{bm}% bold math
\usepackage{lipsum}
\usepackage{hyperref}
\usepackage{xcolor}
\usepackage{soul}
\definecolor{mylinkcolor}{rgb}{0,0,0.8} % set link colour here as red, green, blue.
\hypersetup{unicode=true,%
  bookmarksnumbered=false,bookmarksopen=false,bookmarksopenlevel=1, %
  breaklinks=true,pdfborder={0 0 0},colorlinks=true}%
\hypersetup{%
  anchorcolor=mylinkcolor,citecolor=mylinkcolor, %
  filecolor=mylinkcolor,linkcolor=mylinkcolor, %
  menucolor=mylinkcolor,runcolor=mylinkcolor, %
  urlcolor=mylinkcolor}

\newcommand{\ket}[1]{| #1 \rangle}
\newcommand{\bra}[1]{\langle #1 |}

\newcommand\bin{\mathrm{bin}}
\newcommand{\dd}{\mathrm{d}} % differential
\DeclareMathOperator{\e}{e}

\newcommand{\comment}[1]{}
\def\M{\ensuremath\mathcal}
\def\B{\ensuremath\mathbf}

\newcommand\blfootnote[1]{%
  \begingroup
  \renewcommand\thefootnote{}\footnote{#1}%
  \addtocounter{footnote}{-1}%
  \endgroup
}

\newtheorem{lemma}{Lemma}
\newtheorem{theorem}{Theorem}

\theoremstyle{definition}
\newtheorem{definition}{Definition}

\newtheorem*{protocol}{Protocol}

\AtBeginDocument{%
    \newwrite\bibnotes
    \def\bibnotesext{Notes.bib}
    \immediate\openout\bibnotes=\jobname\bibnotesext
    \immediate\write\bibnotes{@CONTROL{REVTEX41Control}}
    \immediate\write\bibnotes{@CONTROL{%
    apsrev41Control,author="08",editor="1",pages="1",title="0",year="0"}}
     \if@filesw
     \immediate\write\@auxout{\string\citation{apsrev41Control}}%
    \fi
  }%
  
\begin{document}
\vspace{2cm}
\title{Enhancing key rates of QKD protocol by Coincidence Detection}
\begin{abstract}
    In theory, quantum key distribution (QKD) provides unconditional security; however, its practical implementations are susceptible to exploitable vulnerabilities. This investigation tackles the constraints in practical QKD implementations using weak coherent pulses. We improve on the conventional approach of using decoy pulses by integrating it with the coincidence detection (CD) protocol. Additionally, we employ an easy-to-implement technique to compute asymptotic key rates for the protocol secure under specific photon number splitting (PNS) attacks. Furthermore, we have carried out an experimental implementation of the protocol, where we demonstrate that monitoring coincidences in the decoy state protocol leads to enhanced key rates under realistic experimental conditions. 
\end{abstract}
\author{Tanya Sharma$^{1,2 \dagger}$, Rutvij Bhavsar$^{3,4*}$, Jayanth Ramakrishnan$^{1}$, Pooja Chandravanshi$^{1}$, Shashi Prabhakar$^{1}$, Ayan Biswas$^{5}$, and R. P. Singh$^{1**}$}
\affiliation{
$^{1}$ Quantum Science and Technology Laboratory, Physical Research Laboratory, Ahmedabad, India 380009 \\ 
$^{2}$ Indian Institute of Technology, Gandhinagar, India 382355 \\ 
$^{3}$ Department of Mathematics, University of York, Heslington, York, YO10 5DD, United Kingdom \\
$^{4}$ School of Electrical Engineering, Korea Advanced Institute of Science and Technology (KAIST), Daejeon, Republic of Korea \\
$^{5}$ School of Physics, Engineering $\&$ Technology and York Centre for Quantum Technologies, University of York, York, YO10 5FT, United Kingdom
}

\date{\today}
\maketitle
%%%%%%%%%%%%%%%%%%%%%%%%%%%%%%%%%%%%%%%%%%%%%%%%%%%%%%%%%
%                  ~Introduction~                      %
%%%%%%%%%%%%%%%%%%%%%%%%%%%%%%%%%%%%%%%%%%%%%%%%%%%%%%%%%

\section{\label{sec:Introduction}Introduction}
\blfootnote{$\dagger $\href{mailto:tanyabindu@gmail.com}{tanyabindu@gmail.com} } 
\blfootnote{$*$ \href{mailto:rutvij@kaist.ac.kr}{rutvij@kaist.ac.kr}}
\blfootnote{$**$ \href{mailto:rpsingh@prl.res.in}{rpsingh@prl.res.in}}
QKD offers information-theoretic security, allowing two authenticated parties to exchange information securely through a secret and random key. The first QKD protocol was proposed in 1984 \cite{Bennett1984} followed by many others \cite{E91, BBM92, DPSK, COW, MDI, Lucamarini2018}. The principles of quantum mechanics guarantee security without relying on any assumptions regarding the adversary's capabilities \cite{Lo1999, Shor2000, Mayers2001}. While QKD promises unconditional security, its practical implementations are fraught with numerous vulnerabilities that a potential eavesdropper can exploit \cite{Huttner1994, Brassard2000, Vakhitov2001, Makarov2005, Gisin2006, Lydersen2010, Jain2011, Biswas2021}. Security proofs that take into account device imperfections were introduced in,\cite{Lutkenhaus2000, Gottesman, Inamori2007}. 

The imperfect sources provide one such loophole. In most implementations of prepare and measure protocols, weak coherent pulses (WCPs) are used instead of a single photon source (SPS). These weak coherent pulses follow Poissonian statistics; hence, there is a non-zero probability of multiphoton pulses. An adversary can exploit the presence of multiple photons in a pulse to perform a photon number splitting (PNS) attack \cite{Brassard2000,Lutkenhaus2002} to gain information on the shared key.

Decoy state protocol \cite{Hwang2003} was proposed to overcome a PNS attack. Decoy state protocol involves sending additional pulses with varying intensities alongside the signal pulses. By integrating the work of \cite{Gottesman} and \cite{Hwang2003}, the framework for the decoy-state protocol was built \cite{Lo2005, Wang2005, Ma2005}, which is one of the most widely implemented QKD protocols.

Coincidence monitoring is another alternative strategy to detect a PNS attack. The expected and observed coincidences at the receiver's end are compared for a source emitting weak coherent pulses with known photon statistics and a well-characterized channel. Earlier studies \cite{felix_faint_2001, Biswas2022, Lim:17} have proposed monitoring the coincidences to limit the eavesdropper's information on the key, while some of them also suggest using these coincidences to enhance the key rate. However, \cite{Lutkenhaus2002, PhysRevResearch.5.023065} suggests that an eavesdropper can perform a sophisticated PNS attack, where she can mimic the photon statistics. 

In the present study, we propose to combine the decoy state protocol with coincidence detection to circumvent a large class of such sophisticated PNS attacks. Following the common convention, we refer to the sender as Alice, the receiver as Bob, and the adversary as Eve. We monitor the coincidences not only for the signal but also for the decoy states. Since Eve does not know whether a signal or decoy is sent, it is impossible for Eve to practically mimic the statistics at the receiver's end. This is done in addition to the standard decoy state protocol, adding an extra security layer. 

If coincidence monitoring indicates no PNS attack, we can incorporate contributions from two-photon pulses to achieve higher key rates, considering the information leakage through such pulses during a coherent attack. This approach, derived by Lim et al. \cite{Lim:17}, ensures that the computed key rates remain secure under the assumptions of the characterized source and channel. The integration of decoy states and coincidence detections helps us to strategically leverage their capabilities to our advantage. From now on, we will refer to these decoy pulses as entrapped ones to avoid confusion with the decoy state protocol and introduce the entrapped pulse coincidence detection (EPCD) protocol. This integration of entrapped pulses and coincidence detections allows us to strategically exploit their functionalities.

Computing the key rate of a general QKD protocol is a difficult problem since it requires solving a non-linear optimization problem. The analytical technique in the decoy state protocol \cite{Ma2005} gives a tighter bound on the yields when using many decoy states. However, during practical implementations with, say, a single decoy state, this can lead to pessimistic bounds. We present a simple-to-implement method based on semi-definite programming (SDP) to compute the key rate. Our method casts the optimization problem for computing the key rate into a sequence of converging polynomial optimization problems, each of which can be efficiently solved using well-known numerical techniques. We employ our technique to demonstrate that our proposed protocol outperforms the decoy state protocol when the coincidence monitoring allows us to certify the absence of the PNS attack.

This research article is organized as follows. In Sec. \ref{Sec: Theory}, we discuss the theoretical prerequisite and introduce the protocol. In Section \ref{sec: key rate}, we introduce the method to compute the secret key rate for the protocol. Sec. \ref{sec:Experiment} describes the experimental setup and data analysis method. We discuss the results of our experiment and our implementation of SDP for computing key rates in Sec. \ref{sec:RnD} and conclude this research article in Sec. \ref{sec:conc}.

\begin{figure*}[htp]
    \centering
     \includegraphics[scale = 0.5]{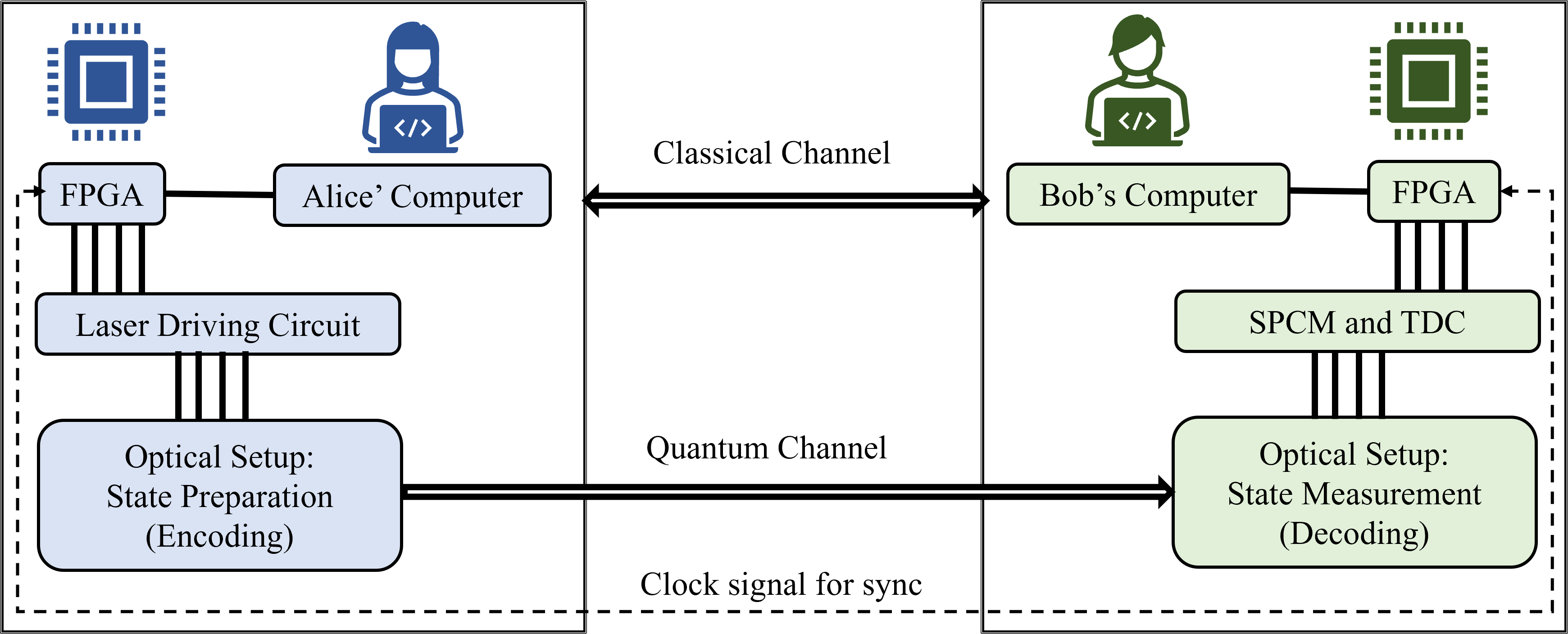}
    \caption{Standard Scheme for Quantum Communication in free space with existing classical systems}
    \label{fig:block_diagram}
\end{figure*}
%%%%%%%%%%%%%%%%%%%%%%%%%%%%%%%%%%%%%%%%%%%%%%%%%%%%%%%%%
%             ~Theoretical Background~                  %
%%%%%%%%%%%%%%%%%%%%%%%%%%%%%%%%%%%%%%%%%%%%%%%%%%%%%%%%%
\section{Theoretical Background}\label{Sec: Theory}
In this section, we first discuss the secure key rates for decoy and CD protocol and then define our EPCD protocol, along with the methods to compute key rates.

\subsection{Secure key rate: Decoy state Protocol}
We employ weak coherent pulses in practical implementations of the BB84 protocol hence, there is a non-zero probability of multiphoton pulses. However, we only consider the contribution from the single-photon pulses in the key rate. The asymptotic secure key rate\cite{Ma2005} is given as:
\begin{eqnarray}\label{eqn: decoy_DW rate}
R \geq \frac{1}{2} \left\{ - Q_{\mu} H_\bin(E_{\mu}) f(E_{\mu}) + Q_1 \left( 1 - H_\bin(e_1) \right)\right\}.
\end{eqnarray}
Here, $\mu$ is the mean photon number of the signal pulses, $Q_\mu$ is the overall gain of signal states, $E_\mu$ is the overall quantum bit error (QBER), $Q_1$ is the gain of single-photon states, $e_1$ is the error rate of single photon states and $H_\bin(e)$ is the binary entropy.

$Q_\mu$ and $E_\mu$ are the experimental parameters, however $Q_1$ and $e_1$ needs estimation. The traditional analytical methods in \cite{Ma2005} compute lower bounds on the key rate by establishing bounds on $Q_1$ and $e_1$.

\subsection{Secure key rate: Coincidence Detection Protocol}

We can detect a PNS attack by monitoring the coincidences, given a well characterized source and channel. This monitoring allows us to determine whether the photon statistics have been altered or remain unchanged \cite{Biswas2022}. If the statistics are unchanged, we are assured that no PNS attack has been performed. However, Eve can still perform the collective and unambiguous state discrimination (USD) attack \cite{USD}.

Let us consider the collective attacks without the USD attack. Acoording to \cite{Lim:17} the attack results in the maximum mutual information between Alice and Eve, $I(A; E)_i$, which can be expressed as:
\begin{equation}
I(A; E)_i = H_\bin\left( \frac{1 + \cos^i c}{2} \right),
\end{equation}
where $\cos c = 1 - 2e_i$. 
The key rate is given as;
\begin{eqnarray}
    R \geq \frac{1}{2} \Bigg\{ 
    && -Q_{\mu} f(E_{\mu}) H_{\mathrm{bin}}(E_{\mu}) \nonumber + Q_1 \left[ 1 - H_{\mathrm{bin}}(e_1) \right] \nonumber \\
    && + \sum_{i=2}^{\infty} Q_i \left( 1 - I(A; E)_i \right) 
    \Bigg\}
\end{eqnarray}

Here, $i$ represents the state containing $i$ number of photons. $Q_i$ and $e_i$ are the gain and error rate of this $i$-photon state, respectively. $H_\bin(e)$ is the binary entropy.

Now, we consider when Eve performs collective and USD attacks. If Eve employs the USD attack, it will certainly fail on two-photon pulses but might have a non-zero success probability with three or more photon pulses \cite{USD}. Hence, we consider the contributions due to just single and two-photon states.
\begin{eqnarray}\label{eqn:DW_rate}
R &\geq& \frac{1}{2} \big\{ - Q_{\mu} H_\bin(E_{\mu}) f(E_{\mu}) \nonumber + Q_1 \left( 1 - \Phi(2 e_1 - 1) \right) \nonumber \\
&& + Q_2 \left( 1 - \Phi((2 e_2 - 1)^2) \right) \big\}.
\end{eqnarray}
Where $\phi(x)$ is defined as follows:
\begin{eqnarray}
    \Phi(x) &:=& H_\bin \left(\frac{1}{2} + \frac{x}{2} \right).
\end{eqnarray}

Here, $Q_2$ and $e_2$ denote the gains of two photos and the error rates, respectively. When calculating the secure key rates for the protocol, the values of gains $Q_1$, $Q_2$, and the error rates $e_1$,$e_2$, are unknown. In a typical decoy state protocol that employs an infinite number of decoy states, accurate estimation of values is achievable. However, this method requires significant resources and is practically unattainable. Ma et al. \cite{Ma2005} suggested using approximations with a single decoy state, but these approximations tend to underestimate the key rates. Here, we propose leveraging the current setup by utilizing coincidences alongside optimization techniques to establish tighter bounds on the key rates. In Sec. \ref{sec: key rate}, we elaborate on our methodology for computing key rates.

\subsection{EPCD Protocol}\label{sec: protocol}
We now define our protocol of using entrapped pulses along with the two-photon coincidences, i.e. entrapped pulse coincidence detection (EPCD) protocol.

\begin{protocol}[Entrapped Pulse Coincidence Detection Protocol (EPCD)]\label{prot: EPCD protocol} In this protocol, we present a comprehensive approach to utilizing entrapped pulses, along with monitoring coincidences. Alice sends a signal characterized by an average photon number $\nu_0$, along with $K$ entrapped pulses characterized by average photon numbers $\nu_1, \nu_2, \cdots, \nu_K$. For a sequence of $n$ pulses, each identified by an index $i$, the subsequent actions are iterated for each pulse.

\begin{enumerate}
   \item \label{step: one} \textbf{State preparation (Alice's lab):} Alice generates random numbers $\M{D}_{i} \in \{ 0 , 1 , \cdots K \}$, $\M{X}_i \in \{ 0 , 1 \}$ and $\M{A}_i \in \{ 0 , 1\}$. Where, 
   \begin{enumerate}
       \item $\M{D}_i$ determines whether a signal ($\nu_0$) or an entrapped pulse ($\nu_1 , \nu_2 , \cdots \nu_K$) is transmitted. If $\M{D}= d$ then mean photon number is $\nu_d$
       \item $\M{X}_i$ determines the basis in which pulse $i$ is encoded. If $\M{X}_i = 0(1)$, she encodes in Standard (Hadamard) basis.
       \item $\M{A}_i$ is the bit value encoded for pulse $i$.
   \end{enumerate}
    \begin{figure}[htp]
        \centering
        \includegraphics[scale = 0.5]{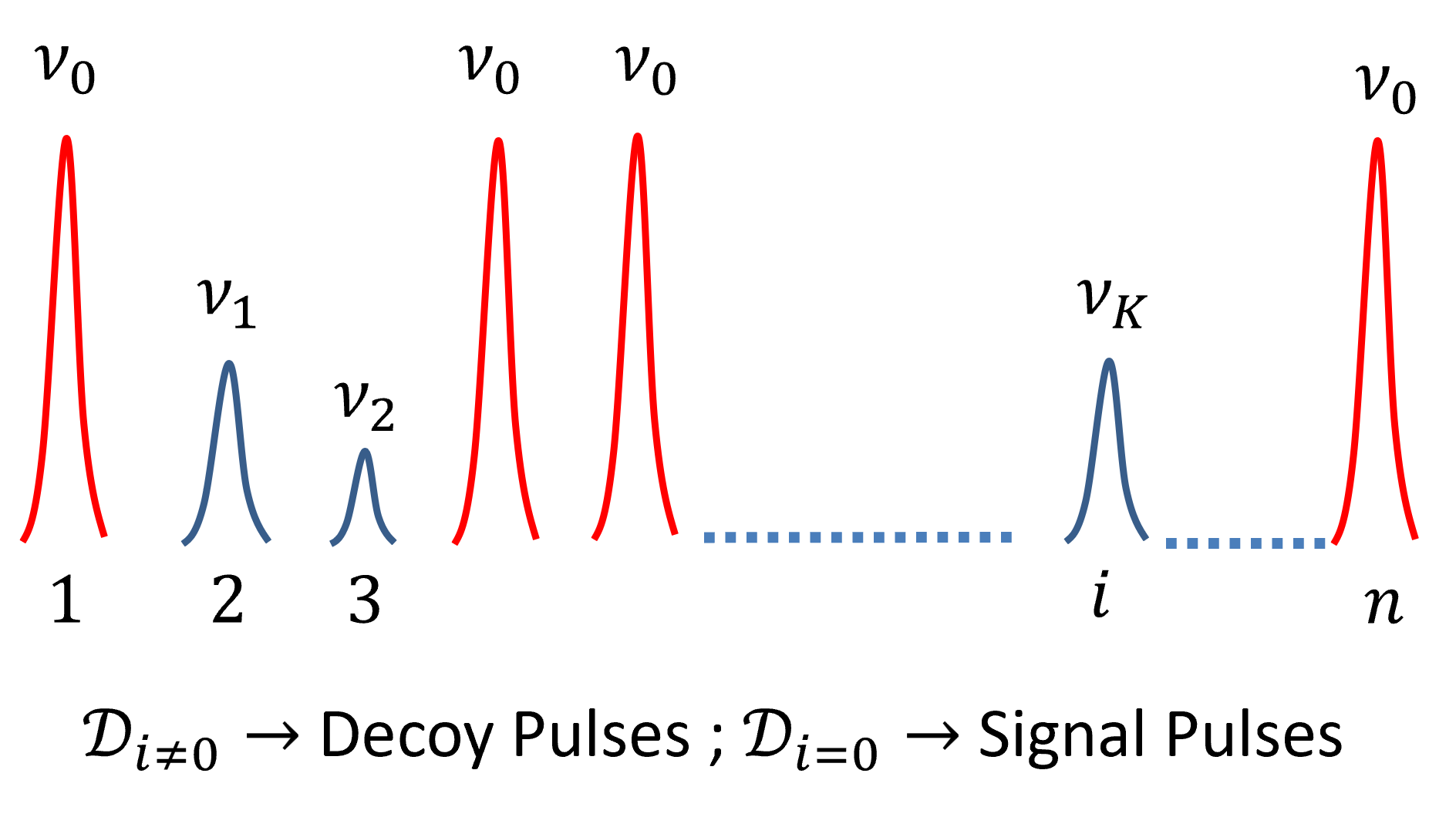}
        \caption{Graphical representation of signal and entrapped pulses randomly transmitted by Alice}
        \label{AL1}
    \end{figure}
   \item \textbf{State transmission:} Alice sends the prepared state to Bob through an quantum channel.
   \item \label{step: Bob} \textbf{State measurement (Bob's lab):} Bob lets the pulse pass through a beam splitter, having two ports $\mathfrak{B}^{(0)}$ and $\mathfrak{B}^{(1)}$. At wing, $\mathfrak{B}^{(0)}$ ($\mathfrak{B}^{(1)}$) Bob performs measurement in the Standard (Hadamard) basis. Let $\M{B}_i^{(j)}$ denote the measured bit value for the pulse $i$ in port $j$.
   \begin{figure}[htp]
        \centering
         \includegraphics[scale = 0.5]{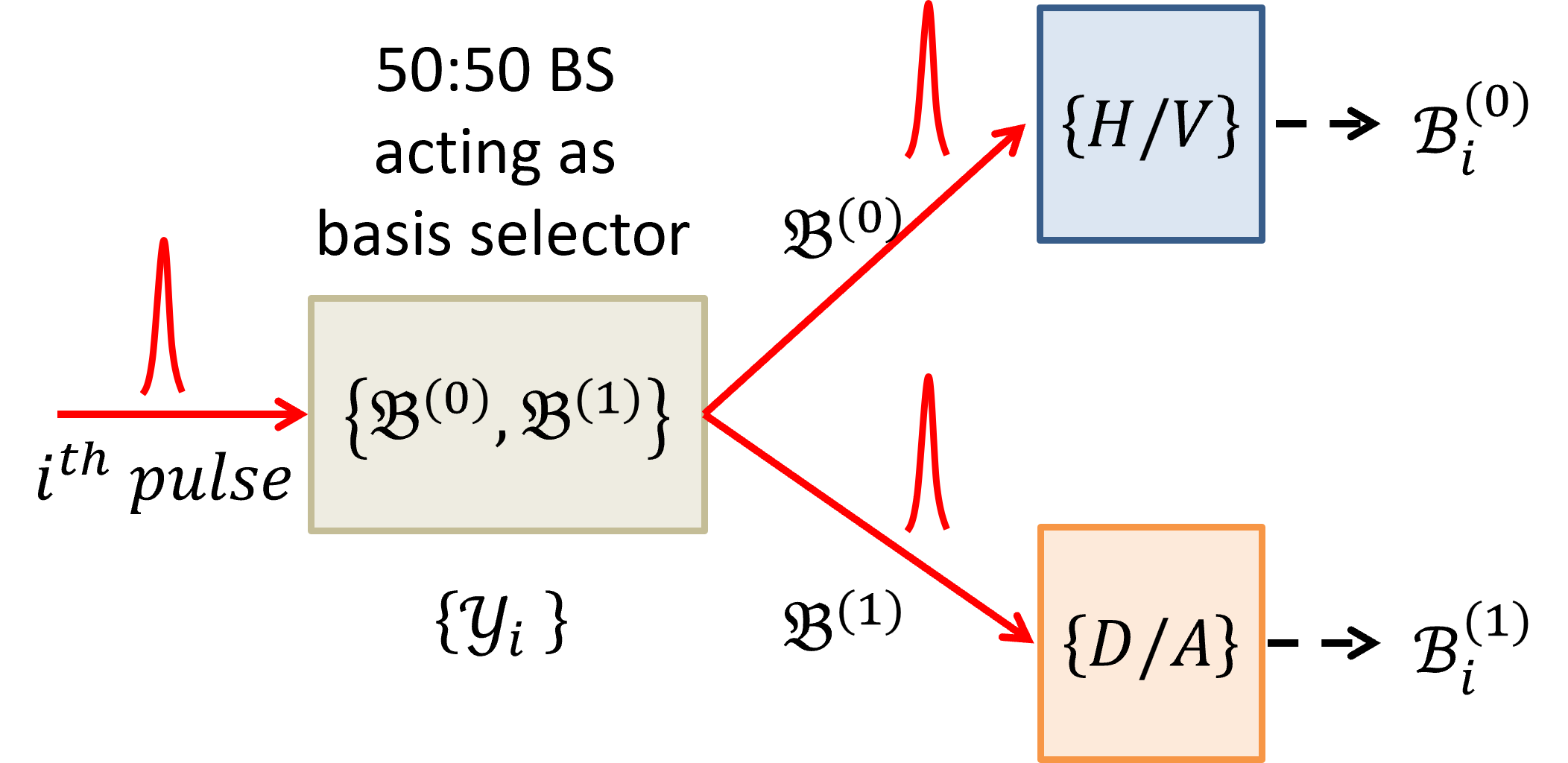}
        \caption{Graphical representation of basis selection and recorded measurements at Bob's end for $i^{th}$ pulse}
        \label{Bb1}
    \end{figure}
   \item Alice and Bob continue to prepare and measure, incrementing the value of $i$ to $i+1$ until $i = n$. Once $i = n$, they proceed to the next step.
   \item Alice publicly discloses the value of her basis $(\M{X}_1 , \M{X}_2 , \cdots , \M{X}_{n}) $.
   \item Let $\M{Y}_{i}$ denote Bob's basis choice and $\M{B}_{i}$ is the measured bit value. No measurement outcome is denoted as $\perp$. 
   \begin{enumerate}
       \item If $\M{B}_{i}^{(j\oplus 1)} = \perp$ and $\M{B}_{i}^{(j)} \in \{ 0 ,1\}$: $\M{Y}_i = j$ , $\M{B}_{i} = \M{B}_{i}^{(j)}$. 
       \item   If $\M{B}_{i}^{(0)} , \M{B}_{i}^{(1)} \in \{ 0 , 1 \}$, then Bob sets $\M{Y}_i = x$ and $\M{B}_i = \M{B}_{i}^{(x)}$ if $\M{X}_i = x$.
       \item Else Bob sets $\M{Y}_i$ randomly and $\M{B}_i = \perp$.
   \end{enumerate}
   \item Bob publicly discloses the values of the basis $(\M{Y}_1, \M{Y}_2, \dots, \M{Y}_n)$. Alice and Bob then discard the rounds in which $\M{X}_i \neq \M{Y}_i$, i.e., they only focus on the rounds from the set $\mathfrak{h} \coloneqq \{ i \in \{1, 2, \dots, n\}: \M{X}_i = \M{Y}_i \}$.
   \item Alice also discloses the entrapped pulses used in each round $(\M{D}_1 , \M{D}_2 , \cdots , \M{D}_{n}) $.
   \item Alice and Bob perform error correction and privacy amplification as in standard BB84 protocol. From any randomly chosen subset $\mathfrak{g} \subset \mathfrak{h}$, with $|\mathfrak{g}|  \ll |\mathfrak{h}|$ they estimate Gains (probability that a signal sent by Alice is received by Bob): 
   \begin{eqnarray}  
       Q_{\nu_d} &:=& \frac{|\{i \in \mathfrak{g}: \M{D}_i = d ,  \M{B}_i \ne \perp \}|}{|\{i \in \mathfrak{g}: \M{D}_i = d \}   | } 
   \end{eqnarray}
   and QBER $\left(E_{\nu_d}\right)$ i.e. the probability that Alice's bit value does not match with Bob's bit value $\left(\M{A}_i \neq \M{B}_i\right)$ whenever they both prepared and measured in same basis $\left(\M{X}_i = \M{Y}_i\right)$):
   \begin{eqnarray} 
       E_{\nu_d} &:=& \frac{|\{i \in \mathfrak{g}: \M{D}_i = d ,  \M{A}_i \neq \M{B}_i \}|}{|\{i \in \mathfrak{g}: \M{D}_i = d  \}|} 
   \end{eqnarray}
\end{enumerate}
\end{protocol}

The above is a generalised description of the protocol; for consistency, we will use $\nu_0 = \mu$ to represent the mean photon number of the signal state. The implementation of the standard decoy state protocol closely follows the procedure outlined in the above protocol, involving one beam splitter and two branches denoted as $\mathfrak{D}^{(0)}$ and $\mathfrak{D}^{(1)}$. While measurements in the standard basis are performed by $\mathfrak{D}^{(0)}$, $\mathfrak{D}^{(1)}$ conducts measurements in the Hadamard basis. The primary distinction lies in the coincidence monitoring, with no significant impact on the experimental feasibility of the protocol. The experimental demonstration of this protocol is discussed in Sec. \ref{sec:Experiment}.

As noted in \cite{felix_faint_2001, Lim:17, Biswas2022}, it has been observed that the PNS attack can be detected by simply measuring the coincidences, even without the presence of entrapped pulses. However, Eve can mimic the photon statistics at Bob's end \cite{Lutkenhaus2002} if we only send the signal without the entrapped pulses. Nevertheless, this type of attack can be effectively countered by employing entrapped pulses. Since the adversary is uncertain whether the transmitted state is an entrapped pulse, she cannot replicate the photon statistics for both the signal and the entrapped pulses.
% However, our protocol may still not detect the most advanced forms of modifications in the PNS attack. In this sophisticated attack, Eve introduces a photon only if she has stored one. However, we argue that implementing such an attack is not practical for Eve with current technology. For Eve to carry out this attack, she would need to first ascertain that she has stored a photon, generate a replacement photon, and then transmit it to Bob. This, particularly with present-day technology, is likely to introduce a noticeable time delay, which Bob can detect. Therefore, it is advisable for Bob to leverage the time delay information to disregard rounds exhibiting significant time disparities.

\section{Computing Key-rate}\label{sec: key rate}
To compute the key rate, we must minimize the key rate over all the strategies that achieve the observed experimental statistics. We will consider this case by case. Let's first address the key rate for coincidence detection given by Eq.\eqref{eqn:DW_rate}. We can substitute, $Q_1 = Y_1 \mu e^{-\mu} $ and $Q_2 =  Y_2 \frac{\mu^2}{2} e^{-\mu}$. Here, $Y_1$ and $Y_2$ are single and two-photon yields, respectively (see appendix \ref{app: ideal protocol}). Eq.\eqref{eqn:DW_rate} now becomes,
\begin{eqnarray}
R \geq && \frac{1}{2} \Big\{ - Q_{\mu} H_\bin(E_{\mu}) f(E_{\mu}) \nonumber \\
&& + Y_1 \mu e^{-\mu} \left( 1 - \Phi(2 e_1 - 1) \right) \nonumber \\
&& + Y_2 \frac{\mu^2}{2} e^{-\mu} \left( 1 - \Phi((2 e_2 - 1)^2) \right) \Big\}.
\label{eqn:DW_rate2}
\end{eqnarray}

\noindent
We know that $Q_\mu$ and $E_\mu$ are experimental parameters hence, the optimisation problem reduces to the last two terms of Eq.\eqref{eqn:DW_rate2}. We take out the common factor of $e^{-\mu}$. Hence, the key rate can be obtained by solving the following optimization problem:

\begin{equation}\label{eqn: basic optimization problem-1}
\begin{aligned}
  r := \min \Big( & Y_1 \mu \left( 1 - \Phi\left(2 e_1 - 1 \right) \right) \\
  & + Y_2 \frac{\mu^2}{2} \left( 1 - \Phi\left( (2 e_2 - 1)^2 \right) \right) \Big) \\
  \text{s.t.} & \quad \forall k: \, Y_k , e_k \in [0 , 1], \\
  & \quad \forall d: \, Q_{\nu_d} e^{\nu_d} = \sum_{k=0}^{\infty} Y_k \frac{\nu_d^k}{k!}, \\
  & \quad \forall d: \, E_{\nu_d} Q_{\nu_d} e^{\nu_d} = \sum_{k=0}^{\infty} e_k Y_k \frac{\nu_d^k}{k!}.
\end{aligned}
\end{equation}

The reason for performing such an optimization is straightforward to understand. The objective function of the optimization problem concerns the key rate $R$ (refer to Eq. \eqref{eqn:DW_rate}). This key rate relies on the $k$ photon yields $Y_{k}$ and $k $ photon error rates $e_{k}$, which are not directly known from experimental statistics. However, what is known are the overall gain $Q_{\mu}$ and the QBER ($E_{\mu}$). 
When we consider just the coincidence detection protocol, the constraints of the optimization problem are related just to the Gain and QBER of the signal. However, when we include the entrapped states, we get an extra set of constraints on the optimization problem from the Gain and QBER of the signal and the decoy states. The more decoy states, the better; however, for practical reasons, we have constrained ourselves to just a single decoy state. 

\noindent
Now we consider the cases when coincidences are not considered, i.e. the cases of BB84 and decoy state protocol, the key rate to be estimated is given by,
\begin{eqnarray}
R \geq && \frac{1}{2} \Big\{ - Q_{\mu} H_\bin(E_{\mu}) f(E_{\mu}) \nonumber \\
&& + Y_1 \mu e^{-\mu} \left( 1 - \Phi(2 e_1 - 1) \right) \Big\}.
\label{eqn:DW_rate_bb84decoy}
\end{eqnarray}

In this case, the optimization problem for computing the key rate differs from Eq. \eqref{eqn: basic optimization problem-1} only in the objective function. Specifically, it does not include contributions from the two-photon terms $Y_2$ and $e_2$. Hence, the optimisation problem is:
\begin{equation}\label{eqn: basic optimization problem-2}
\begin{aligned}
  r :=   \min&  \big( Y_1 \mu \left( 1 - \Phi\left(2 e_1 -1 \right) \right)\big)\\ 
     \mathrm{s.t.}       &  \forall k: \quad Y_{k} , e_{k} \in [0 , 1]  \\
      & \forall d: \quad Q_{\nu_d} e^{\nu_d}  =   \sum_{k = 0}^{\infty} Y_{k} \frac{\nu_d^{k}}{k!}  \\ 
      & \forall d: \quad E_{\nu_d} Q_{\nu_d} e^{\nu_d} =  \sum_{k = 0}^{\infty} e_{k} Y_{k} \frac{\nu_d^{k}}{k!} 
\end{aligned}
\end{equation}
It is to be noted that for the non-decoy case $d=0$, there is just a single intensity, $\nu_0 = \mu$. For the decoy state, the number of constraints depends on the number of decoy states employed.

\noindent
Given that no assumptions can be made about the Yields $Y_k \in [0, 1]$ and error rates $e_k \in [0, 1]$, we acknowledge the possibility of a potential adversary to manipulate them freely. The adversary is, in principle, unrestricted in choosing any values for these parameters, provided that the chosen values align with the observed experimental statistics. This fact is captured using the constraints in the optimization problem, limiting the choice available to the adversary. To account for the strategy that is most advantageous to the adversary within these constraints, we minimize the rate. 

\noindent
In the literature, Ma et al. \cite{Ma2005} conducted the first comprehensive exploration of lower bounds on the asymptotic key rates for the decoy state protocol. For the case with only one decoy, traditional analytical methods compute lower bounds on the key rate by first establishing bounds on $Y_1$ and $e_1$:
\begin{eqnarray}
    Y_0 \leq Y_0^U &:=& \frac{E_{\nu}Q_{\nu}e^{\nu}}{e_0}, \label{eq:y0} \\ 
    Y_1 \geq Y^L_1 &:=& \frac{\mu}{\mu\nu - \nu^2} \left(Q_{\nu} e^{\nu} - Q_{\mu} e^{\mu} \left(\frac{\nu^2}{\mu^2}\right) \right. \nonumber \\
     &&\left. - Y_0^U \left(\frac{\mu^2 - \nu^2}{\mu^2}\right)\right), \label{eq:y1}\\
    e_1 \leq e_1^U &:=& \frac{E_{\nu} Q_{\nu} e^{\nu}}{Y_1^L \nu} \label{eq:e1}
\end{eqnarray}

\noindent 
Exploiting the monotonicity properties of binary entropy, specifically that $H_{\bin}(x)$ is increasing for all $x \in [0 , 1/2]$, and the fact that $(1 - H_{\bin}(x)) > 0$, a lower bound on the secure key rate \eqref{eqn:DW_rate_bb84decoy} is given by:
\begin{eqnarray}\label{eq:R_low}
    R \geq && \frac{1}{2} \Big\{ - Q_{\mu} f(E_{\mu})H_{\bin}(E_{\mu}) \\
    && +  Y_{1}^{L}\mu e^{-\mu} \left( 1 -  H_{\bin}(e_{1}^{U}) \right) \Big\}. 
\end{eqnarray}
An immediate approach to computing lower bounds on the key rate for the protocol with coincidence detection involves employing similar analytic techniques to establish upper and lower bounds, denoted as $Y_2^{L}$ and $e_2^{U}$, on $Y_2$ and $e_2$ respectively, that are compatible with the constraints. These bounds will yield tight results on the key rate in scenarios where many decoy states are sent. However, this process is tedious, and tight bounds may not be easily found. 

For our protocol, aiming for experimental practicality, we use a single decoy to derive the key rate. The results obtained using the aforementioned analytical technique may lead to loose bounds. We seek tight bounds on the key rate when only a few decoys are sent. Numerical methods, which rely on linear programming tools and are easy to implement, can be employed to find such bounds (one approach is shown in Appendix \ref{app: Speeding up the optimization problem}). Furthermore, these bounds yield tighter results on the key rate when more decoy states are sent. The next section introduces a simple-to-implement method to iteratively derive bounds on the key rate that converge to the asymptotic key rate from below.

\subsection{Solving the optimization problem}
Two primary challenges emerge when addressing optimization problem \eqref{eqn: basic optimization problem-1}. First, the problem is non-convex due to the non-convex nature of both the constraints and the objective function in \eqref{eqn: basic optimization problem-1}.   General optimization problems are notoriously difficult to solve due to their abstract nature unless they fit specific classes of convex optimization problems. As a result, the optimization problem cannot be solved directly using conventional techniques. Second, optimization problem \eqref{eqn: basic optimization problem-1} involves infinitely many free parameters, namely, $Y_{k}$ and $e_{k}$. The presence of these infinitely many parameters adds further complexity to the problem.

The challenge of dealing with infinitely many variables can be addressed straightforwardly by recognizing that the contribution of $Y_k$ and $e_{k}$ when $k \gg 1$ is negligible to the sums $\sum_{k} (\nu_{d}^{k}/k!) Y_k$ and $\sum_{k} (\nu_{d}^{k}/k!) Y_{k} e_k$. Consequently, we relax the constraints of \eqref{eqn: basic optimization problem-1} by truncating the infinite sums $\sum_{k} (\nu_d^{k}/k!) Y_k$ to involve sums over a finite number of variables. This comes, however, with a small penalty that depends on the number of terms kept in the sum. We formally implement this truncation in Appendix \ref{app: finite optimization problem}.

We begin by discussing our general technique for addressing certain simple non-polynomial optimization problems, drawing inspiration from~\cite{BhRC2023, bhavsar2023improvements}. This method involves deriving lower bounds for optimization problems through the partitioning of the parameter space into smaller sub-spaces. Formally, this partition (or grid), denoted as $\mathcal{P}$, divides the space into multiple sub-domains $\{\mathcal{C}_{i}\}_{i}$. For each sub-domain $\mathcal{C}_i$, we formulate a new optimization problem (or sub-problem) by constraining the parameters to that sub-domain. Each sub-problem can be lower-bounded by a polynomial optimization problem using simple methods such as Taylor's theorem, with only a minimal loss of tightness. Obtaining lower bounds on the resulting sub-problems is achieved by computing the Semidefinite Programming (SDP) relaxations of the polynomial optimization problem (in practice, this is done using software tools like NCPOL2SDPA \cite{NCPOL2SDPA}). A reliable lower bound on the optimization problem is then determined by identifying the lowest computed value (see Lemma \ref{lem: general_technique}, Appendix\,\ref{app: partitioning the domain}). Detailed information on partitioning and lower-bounding the sub-problems generated by any given partition is formally done in Appendix\,\ref{app: Algo}. 

The solutions to the optimization problems can be improved by refining the partition so that the sub-domains have smaller dimensions. Typically, if the partitioning is done over the parameter space of $P$ independent parameters, then enhancing tightness by a factor of $\beta$ would require computing $\beta^{P}$ times more sub-problems.  Consequently, this technique is particularly applicable and effective when partitioning is applied to a relatively smaller number of parameters.  

Now let us return to the optimization problem at hand. There are only two non-polynomial terms $\Phi(2e_1 - 1)$ and $\Phi((2e_2 - 1)^2)$ in the optimization problem. Furthermore, only two parameters, $e_1$ and $e_2$, contribute to the non-polynomial terms in the problem. As $e_1$ and $e_2$ are both constrained within the range of $[0, 1]$, we construct a partition $\M{P}$ of the set $[0, 1] \times [0, 1]$ , generating rectangular sub-domains $\M{C}_{i}$. Leveraging the properties of the function $\Phi(x)$, we find the constants $\xi^{\max}_{1,i}$ and $\xi^{\max}_{2,i}$ that (tightly) lower bound the functions $\Phi(2e_1 - 1)$ and $\Phi((2e_2 - 1)^2)$ respectively in the sub-domain $\M{C}_{i}$\footnote{To minimize the objective function, a lower-bound on $ - \Phi((2e_{k} - 1)^{k})$ is needed. }(See lemma \ref{lem: lowr_bnd}, Appendix\, \ref{app: Algo}). We then use these tight bounds to lower-bound the objective function as follows
% \begin{eqnarray}
%     \sum_{k \in \{1 , 2\}} Y_k \frac{\mu^{k}}{k!}(1 - \Phi((2e_k - 1)^{k})) \geq \sum_{k \in \{1 , 2\}}Y_k \frac{\mu^{k}}{k!}\cdot (1 - \xi^{\max}_{k,i}).
%   \end{eqnarray}

\begin{eqnarray}
    &&{}\!\!\!\!\!\!\!\!\sum_{k \in \{1 , 2\}} Y_k \frac{\mu^{k}}{k!}(1 - \Phi((2e_k - 1)^{k})) \nonumber \\
    &&\geq \sum_{k \in \{1 , 2\}}Y_k \frac{\mu^{k}}{k!}\cdot (1 - \xi^{\max}_{k,i}).
\end{eqnarray}
The loss in tightness when bounding the objective functions by such constants $\xi^{\max}_{1,i}$ and $\xi^{\max}_{2 ,i}$ depend upon the dimensions of the domain $\M{C}_{i}$ as shown in Lemma \ref{lem: lowr_bnd}, Appendix\,\ref{app: Algo}. Thus, the loss of tightness can be made arbitrarily small by refining the partition. However, making a finer partition comes at the expense of higher computation time, as more and more sub-problems need to be numerically solved. We give a formal proof our technique in Appendix \ref{app: Algo}.

%All the results lead to the following final result, which forms the basis of the algorithm for computing key rate $r$ (as defined in equation \eqref{eqn: basic optimization problem-1}). % We state it here informally as: \\

%\textbf{Theorem 1}(Informal):\textit{ Consider a partition $\mathcal{P} = \{ \mathcal{C}_{i} \}_i $ of the interval $[0 , 1]\times[0 , 1]$. The minimum achievable value $r \geq \underset{i}{\min}\{ r_{n}(\mathcal{C}_{i} ) \}_{i}$, where each $r_{n}(\mathcal{C}_{i})$ corresponds to the solution of a polynomial optimization problem over $2n$ number of variables (given by equation \eqref{eqn: app_optimization}). Moreover, for any given (arbitrarily small) margin of error $\epsilon > 0$, there exists (a sufficiently large) $n$ and a (sufficiently fine) partition $\mathcal{P}$ such that the difference $|r - \underset{i}{\min} \{ r_{n}(\mathcal{C}_{i}) \}_i|$ is at-most $\epsilon$. }\\

%In essence, using our lower bounds can be determined with arbitrarily high precision by solving multiple polynomial optimization problems and then computing the minimum of all the results obtained. 

Assuming a random background with errors occurring equally likely, we explicitly set the background error rate as $e_0 = 1/2$. To account for the impact of finite statistics on the experimental values $Q_{\mu}$, $Q_{\nu}$, $E_{\mu}$, and $E_{\nu}$, it is imperative to consider uncertainties in their measurement. In Appendix \ref{app: rates in presence of statistical errors}, we slightly modify our technique to accommodate these statistical uncertainties in the computation of rates. It is important to note that the key rates reported here are key rates in the asymptotic limit. However, this analysis only serves to acknowledge the uncertainty associated with experimentally observed values, and the computed rates are still asymptotic key rates compatible with experimentally obtained statistics with relevant uncertainties in observed values. The computation of finite-round statistics can, in principle, be performed using theoretical tools such as the Generalized Entropy Accumulation Theorem \cite{GEAT, GEAT-security}. Such analysis would require more sophisticated analysis and is reserved for future work.

%%%%%%%%%%%%%%%%%%%%%%%%%%%%%%%%%%%%%%%%%%%%%%%%%%%%%%%%%
%                ~Experimental Method~                  %
%%%%%%%%%%%%%%%%%%%%%%%%%%%%%%%%%%%%%%%%%%%%%%%%%%%%%%%%%

\section{\label{sec:Experiment}Experimental Method }

This section outlines the experimental procedures involved in implementing our secure QKD protocol. The experimental setup consists of three main stages: state preparation, transmission, and state measurement.

\begin{figure*}[htp]
    \centering
     \includegraphics[scale=0.22]{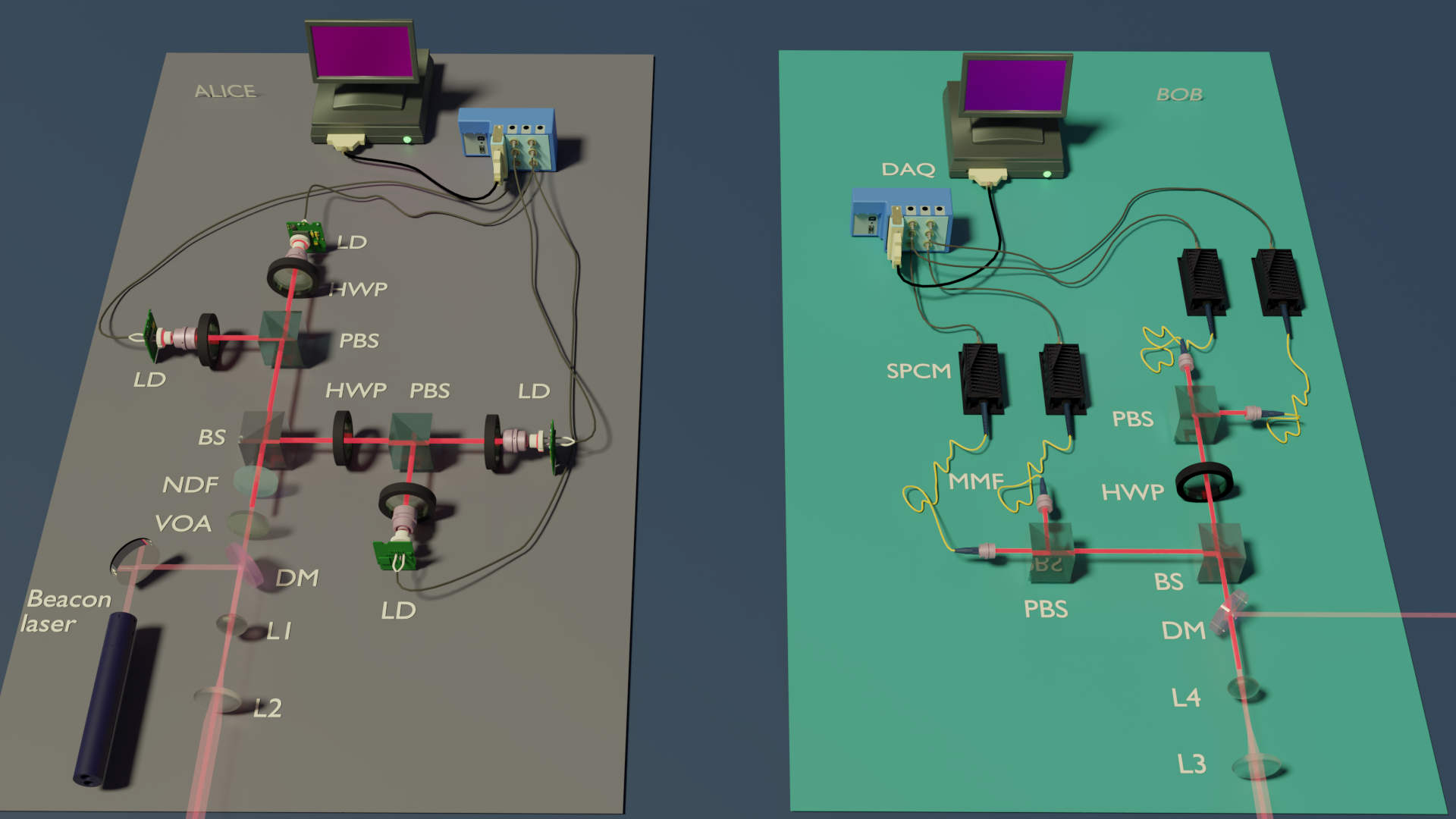}
    \caption{Schematic of the experimental setup. It includes both the optics and electronic components. LD: Laser Diodes, HWP, Half Wave Plate, BS: Beam Splitter, PBS: Polarising Beam Splitter, NDF: Neutral Density Filter, VOA: Variable optical Attenuator, DM: Dichroic Mirror, L: lens, MMF: Multi-Mode Fibre, SPCM: Single Photon Counting Module, DAQ: Data Acquisition system}
    \label{fig:setup}
\end{figure*}

\subsection{State Preparation: Alice}

Alice uses the polarisation degree of freedom of weak coherent pulses to encode the quantum state to be sent. Weak coherent pulses are generated by attenuating the laser pulses using neutral density filters. Our experimental setup Fig.\,\ref{fig:setup} comprises four laser diodes operating at a wavelength of $808$ nm (Thorlabs $L808P010$). An in-house designed laser driving circuit driven by a Field Programmable Gated Array (FPGA) activates the laser diodes, producing pulses at a frequency of $5$ MHz, each with an optical pulse width of $1$ ns. All four lasers are then optically engineered to generate the four polarisations: horizontal (H), vertical (V), diagonal (D) and anti-diagonal (A), using polarizing beam splitters (PBS) and half-wave plates (HWP). The combination of PBS and HWP acts as a polariser as well as aids in the attenuation of the pulse. The FPGA also ensures that the four lasers are randomly triggered, thereby ensuring that the four polarisation states are randomly generated. We have characterized all four sources to account for various side channels arising from discrepancies in various degrees of freedom such as pulse width, wavelength, and power \cite{Biswas2021}. After combining all the pulses on the beam splitter (BS), the signal is passed through one fixed and one variable neutral density filter, which is used to attenuate the pulses further. The variable NDF is rotated to vary the intensity of the signal, measured in terms of mean photon number, characterised by using the method proposed in \cite{sharma_2024}. The decoy pulses are generated using this variable NDF by changing the intensity of the pulses. This enables us to conduct a proof of principle decoy state QKD, using the pulses at two different intensities for generating the signal and the decoy pulses. Thereafter, Alice sends the encoded signal and decoy pulses to Bob via the free space channel.

\subsection{Transmission: The Channel}
The signal transmission to Bob takes place through a free space channel. The communication framework within the Thaltej campus of the Physical Research Laboratory (PRL) in Ahmedabad, Gujarat, India, encompasses two neighbouring buildings. Both the transmitter and receiver stations are situated in adjacent rooms on the rooftop of the first building. A reflector is placed on the second building to direct the signal to Bob. The channel is then characterized for the losses, where we have included the losses due to launching and collecting optics in the channel loss itself. The channel was characterized using the beacon laser of 633 nm, giving the transmittance of $86\%$. The experimental parameters and their values are specified in Table. \ref{table: parametersch5}. We have conducted our experiment in the natural environment rather than a controlled laboratory setting, deliberately incorporating variables such as atmospheric interference and night lighting. This approach ensures a more comprehensive and realistic assessment, enhancing the credibility and applicability of our findings.

\subsection{State Measurement: Bob}
Upon reaching Bob, the signal undergoes a decoding process involving projective measurements onto the four polarisation states. The choice of basis for the measurement is achieved by using a beam splitter (BS). The presence of an HWP $@22.5$ in one of the paths designates the diagonal basis, while the other path represents the rectilinear basis. The signals are detected using single photon counting modules (Excelitas-SPCM-AQRH-14), integrated with a high-performance data acquisition card to record the timestamps of detections across various polarisations. Given the inherent nature of our source as a WCP, a non-zero probability of coincidences exists during the detection process. After a sufficient exchange of quantum states, Alice and Bob proceed to classical postprocessing. 

\begin{table}[h]
\centering
\begin{tabular}{|c|c|}
\hline
\textbf{Parameter} & \textbf{Value} \\
\hline
Channel Transmission & $ 0.86$ \\
\hline
Pulse width & $1$ ns \\
\hline
Detection jitter & $350$ ps \\
\hline
Coincidence window & $2$ ns \\
\hline
Detection Efficiency & $ 0.62$ \\
\hline
Coupling Efficiency & $ 0.87$ \\
\hline
Dark counts & $100$ cps \\
\hline
Background counts & $3000$ cps \\
\hline
\end{tabular}
\caption{The values for channel transmission, pulse width, detection jitter, coincidence window, detector efficiency and coupling efficiency, dark and background counts. The unit cps is for counts per second. The background and dark counts are the worst cases considered.}
\label{table: parametersch5}
\end{table}

\subsection{Data Analysis and Postprocessing}
We recorded the timestamps when each pulse of Alice was emitted in all four polarisations. Bob's recorded data comprises the timestamps for detection clicks in all four detectors. From Bob's time stamps, we estimated the singles and the coincidences. If only one detector clicks for a given time stamp, it is a single detection; otherwise, we record two, three and fourfold coincidences depending on the number of detectors giving a simultaneous click. Alice reveals her chosen basis publicly. For single detections, Bob retains those where his basis is compatible with Alice's basis. In the case of coincidence detection, Bob is only interested in two-fold coincidences, specifically those occurring in opposite basis. In such instances, Bob selects measurements aligned with Alice's basis choice. These steps correspond to Steps 5-8 outlined in the protocol detailed in Sec. \ref{sec: protocol}. The refined bits from this process form the sifted key, a portion of which is used to assess the gain and quantum bit error rate (QBER). This gain and QBER are then further used to estimate the bounds on the key rates of different protocols.

\begin{figure}[htp]
    \centering
    \includegraphics[scale=0.5]{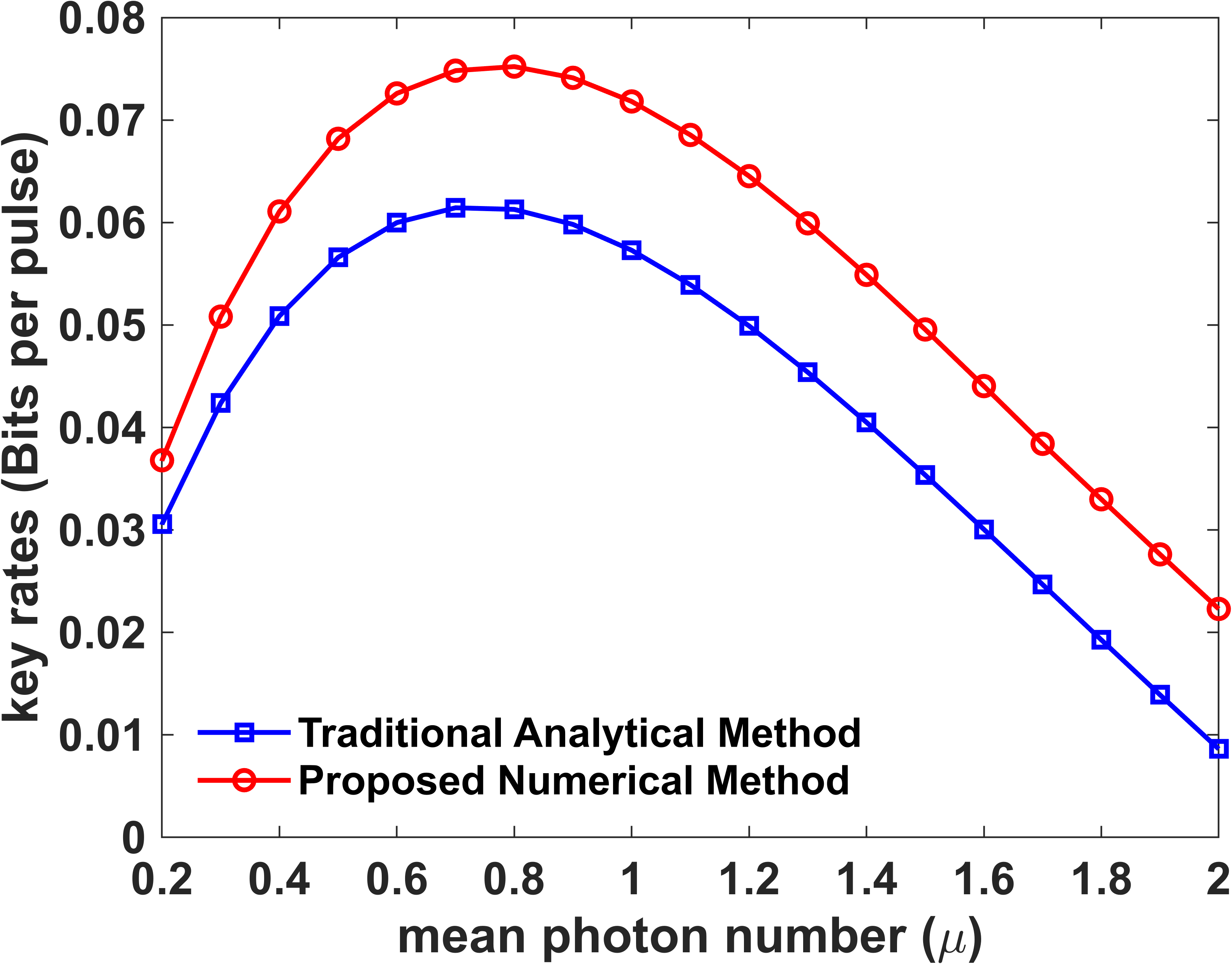}
    \caption{Key rates (bits per pulse) as a function of mean photon number ($\mu$) for two cases: (i) traditional analytical technique and (ii) proposed numerical technique using simulated results}
    \label{fig:comparison_mu}
\end{figure}
\begin{figure}[htp]
    \centering
    \includegraphics[scale=0.5]{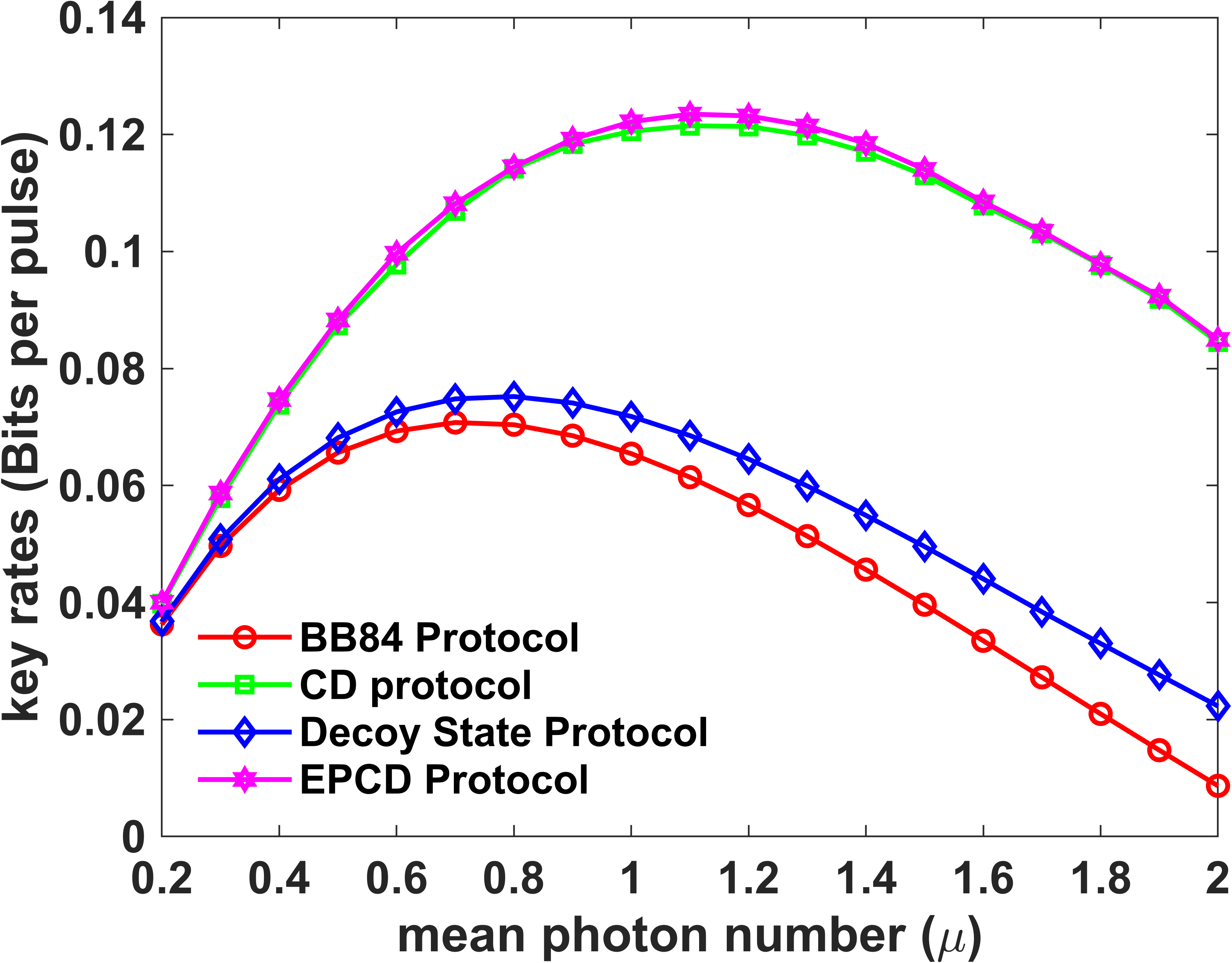}
    \caption{Simulated key rates (bits per pulse) as a function of mean photon number ($\mu$) for four protocols: (i) BB84 Protocol, (ii) CD protocol, (iii) Decoy state protocol and (iv) EPCD protocol.}
    \label{fig:sim_mu}
\end{figure}
\begin{figure}[htp]
    \centering
    \includegraphics[scale=0.5]{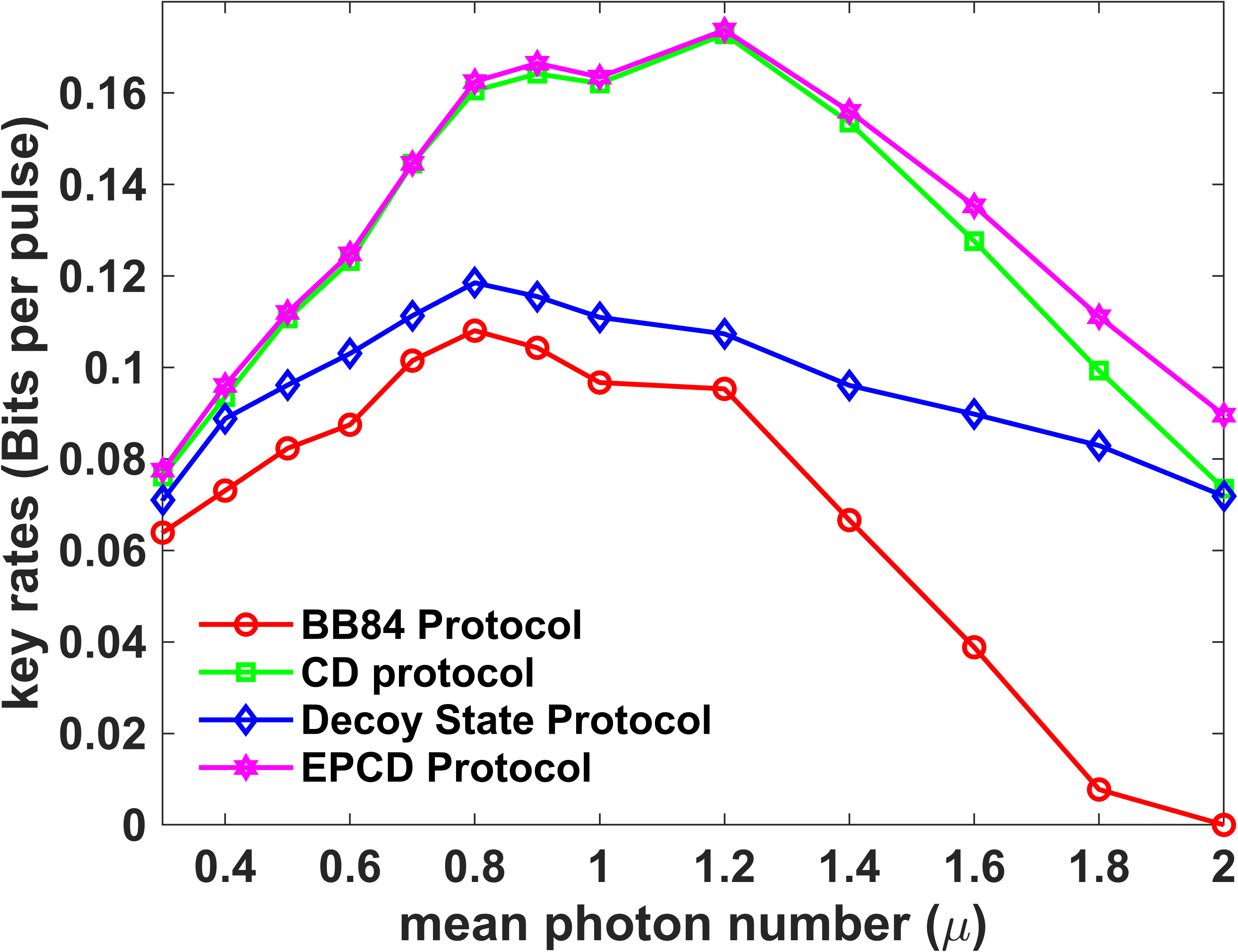}
    \caption{Experimental secure key rates (bits per pulse) as a function of mean photon number ($\mu$) for four protocols: (i) BB84 Protocol, (ii) CD protocol, (iii) Decoy state protocol and (iv) EPCD protocol.
    Lines drawn are for the aid of the eye.}
    \label{fig:exp_mu}
\end{figure}
\begin{figure}[htp]
    \centering
    \includegraphics[scale=0.5]{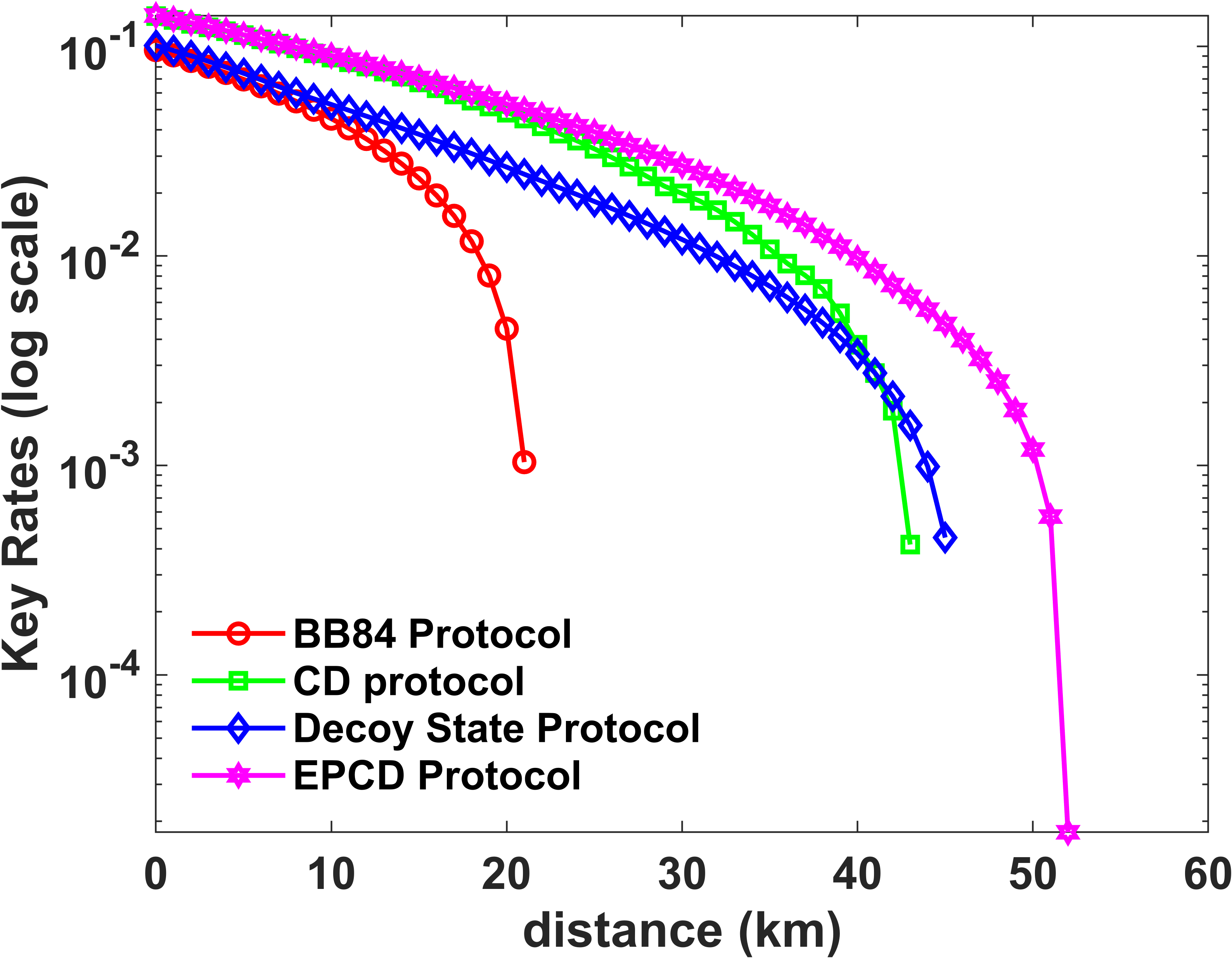}
    \caption{Key rates (bits per pulse) as a function of distance (in km) for four protocols: (i) BB84 Protocol, (ii) CD protocol, (iii) Decoy state protocol and (iv) EPCD protocol. The plots are in log scale}
    \label{fig:sim_d}
\end{figure}

%%%%%%%%%%%%%%%%%%%%%%%%%%%%%%%%%%%%%%%%%%%%%%%%%%%%%%%%%
%              ~Results and Discussions~                %
%%%%%%%%%%%%%%%%%%%%%%%%%%%%%%%%%%%%%%%%%%%%%%%%%%%%%%%%%
\section{\label{sec:RnD}Results and discussion}
Increasing the key rate in realistic scenarios remains a challenge as alternative approaches typically demand additional hardware components or intricate optimization techniques for key extraction. We proposed a solution minimizing the resource requirement on both the software and hardware side. We performed a comparative study of key rates across different protocols. We then evaluated our key rate optimization method against existing ones, with detailed results presented in subsequent paragraphs.

We compare the traditional Decoy state method with the proposed SDP-based method in calculating lower bounds on the secure key rates given in Eq\, \eqref{eq:R_low}. This comparison is performed for various values of the mean photon number as shown in Figure.\,(\ref{fig:comparison_mu}), which is done by performing a simulation of the decoy-based BB84 protocol for various mean photon numbers. In the simulation, the mean photon number of the decoy is kept constant with $\nu=0.1$. It is seen that our proposed method gives tighter lower bounds than the traditional analytic method by $27 \%$ at the optimal mean photon number ($\mu =0.8$).

We employ our SDP-based method to analyse the key rates for the four protocols viz. BB84, decoy state protocol, coincidence detection (CD) protocol, and entrapped pulse coincidence detection (EPCD) protocol. We use the SDP technique to find the bounds on the yields and errors to compute the bounds on key rates using Eqs.\,\eqref{eq:R_low} and \eqref{eqn:DW_rate2}. The respective key rates are plotted as a function of the mean photon number ($\mu$) in Figure.\,(\ref{fig:sim_mu}). It is established that when coincidences are taken into account, the key rate is enhanced by a substantial amount. The coincidence detection protocol helps to achieve a higher optimal mean photon number, i.e. $\mu = 1.1$, as compared to using the traditional decoy-state method without coincidences, i.e. $\mu = 0.8$. Comparing the optimal key rates of decoy-state protocol and EPCD protocol, we note an improvement of $\approx 64\%$ corresponding to higher achievable distances.

Using the experimental setup described in the previous Section, the four protocols are executed. The protocols are systematically repeated for several signal intensities, each corresponding to distinct mean photon numbers. The experimentally acquired gain ($Q_\mu$) and QBER ($E_\mu$) are then used to estimate the optimal bounds on the secure key rate as discussed in Section.\,\ref{sec: key rate}. We have used the timestamps for recording the coincidences in addition to the single detections. The data were taken at various intensities of the source. These two cases enabled us to study four variants for our QKD protocol: Standard BB84, decoy state protocol, CD protocol, and EPCD protocol. Optimal key rates were calculated for all four cases at different values of mean photon number ($\mu$). The key rates for all the cases as a function of the mean photon number can be seen in Figure.\,(\ref{fig:exp_mu}). It is also observed that our experiment reproduces the theoretical predictions of improvement in secure key rates. We obtained a maximum key rate of $0.17$ per pulse for the EPCD protocol.

We have simulated the expected secure key rate for the four protocols as a function of distance over fiber with a loss of 0.2 dB/km. The simulated key rates are plotted in Figure.\,(\ref{fig:sim_d}). In this simulation we have considered the mean photon number of the signal and decoy to be $\mu=0.9$ and $\nu=0.1$, respectively. It is evident that the EPCD protocol consistently gives higher key rates than other protocols across varying distances, showcasing its effectiveness in extending communication ranges. This consistent performance highlights the EPCD protocol's reliability and strength in addressing challenges related to communication distance. 
%%%%%%%%%%%%%%%%%%%%%%%%%%%%%%%%%%%%%%%%%%%%%%%%%%%%%%%%%
%                     ~Conclusion~                      %
%%%%%%%%%%%%%%%%%%%%%%%%%%%%%%%%%%%%%%%%%%%%%%%%%%%%%%%%%
\section{ \label{sec:conc}Conclusion}

In this work, we have proposed the entrapped pulse coincidence detection (EPCD) protocol, where we have integrated two approaches to harness their collective strength. Monitoring coincidences of the signal as well as entrapped pulses allows us to detect the most practical, sophisticated PNS attacks. With an assurance of no PNS attack, we showed that including contributions from two-photon statistics, that we could achieve higher key rates.

A novel method for optimizing key rates using Semi-Definite Programming (SDP) is introduced, providing tight bounds on asymptotic key rates for the protocols discussed in this paper. In practical terms, this method yields reasonable bounds on key rates within a few minutes when executed on a personal computer. Moreover, the method is readily parallelizable, ensuring adaptability for scenarios where computed bounds may need further tightening, although we have not encountered such a situation in our experience thus far. This method has been used to compute the asymptotic key rates for the field implementation of our protocol. The obtained results illustrate that employing the proposed protocol substantially enhances the asymptotic key rates. We plan to conduct a future study on the finite-size analysis of our proposed protocol.
%%%%%%%%%%%%%%%%%%%%%%%%%%%%%%%%%%%%%%%%%%%%%%%%%%%%%%%%%
%                  ~Acknowledgement~                    %
%%%%%%%%%%%%%%%%%%%%%%%%%%%%%%%%%%%%%%%%%%%%%%%%%%%%%%%%%
\section*{Acknowledgement}
The authors acknowledge the partial funding support from DST through the QuST program. The authors are also thankful to group members of the QST lab for their valuable input, with a special thanks to Dr. Satyajeet Patil and Mr. Vardaan Mongia. RB acknowledges support from Quantum Communications Hub of the UK Engineering and Physical Sciences Research Council (EPSRC). AB thanks Prof. Marco Lucamarini and Dr. Rupesh Kumar for insightful discussions. 
%%%%%%%%%%%%%%%%%%%%%%%%%%%%%%%%%%%%%%%%%%%%%%%%%%%%%%%%%
%                    ~Disclosure~                       %
%%%%%%%%%%%%%%%%%%%%%%%%%%%%%%%%%%%%%%%%%%%%%%%%%%%%%%%%%
\section*{Disclosure}
The authors declare no conflicts of interest.
%%%%%%%%%%%%%%%%%%%%%%%%%%%%%%%%%%%%%%%%%%%%%%%%%%%%%%%%%%
          %%% Appendix      
%%%%%%%%%%%%%%%%%%%%%%%%%%%%%%%%%%%%%%%%%%%%%%%%%%%%%%%%%%

\appendix 
\section{Yield, Gain and QBER}\label{app: ideal protocol}

In this section, we discuss the definitions of yields and error rates for completeness. Furthermore, following the work of Ma et al. \cite{Ma2005} we state the expected values of these yields and error rates in an "ideal protocol", in which no eavesdropping is performed, and errors arise naturally due to lossy channels. 

In our protocol, it is assumed that Alice prepares the signal state using phase-randomised weak coherent pulses given by: 
\begin{eqnarray}
    \hat{\rho} &=& \frac{1}{2\pi}\int_{0}^{2\pi} \ket{\mu e^{i \theta}}\bra{\mu e^{i \theta} } d\theta \nonumber \\
    &=& \sum_{n=0}^{\infty}\frac{\mu^n}{n !} \mathrm{e}^{-\mu} \ket{n}\bra{n} \label{eq:coh_state},
\end{eqnarray}
where  $\ket{\mu e^{i \theta}}$ is the coherent state of the electromagnetic field \cite{fox_quantum}. It is evident from Eq.\,\eqref{eq:coh_state}, that the photon number $N$ for the weak coherent pulses, follow the poison distribution with probability $\mathbb{P}(N = n)$ given by Eq.\eqref{eq:probdist}.
\begin{equation}\label{eq:probdist}
    \mathbb{P}(N = n)=\frac{\mu^n}{n !} \mathrm{e}^{-\mu}.
\end{equation}

We adopt identical notations and formalism as presented in \cite{Ma2005} for congruence. Let $\eta$ be the overall efficiency for detection by Bob of a single photon sent by Alice. In an ideal protocol, the efficiency of detection of a $k$ photon pulse is given as:

\begin{equation}\label{eq:7}
  \eta_k= 1- (1-\eta)^k \qquad \mathrm{for} \ k \in \mathbb{N}
\end{equation}

\textbf{Yield:} Yield $Y_k$ is the conditional probability of a detection event at Bob's side. The yield $Y_k$ includes the probability of detection of signal photons $\eta_k$ and the background events $Y_0$. We assume that the background counts and the signal are independent of each other. Then for an ideal protocol,
\begin{equation}\label{eq:8}
\begin{aligned}
    Y_k &=Y_0 + \eta_k - Y_0\eta_k 
\end{aligned}    
\end{equation}

\textbf{Gain:} Gain $Q_\mu$ is the probability with which a signal sent by Alice is detected by Bob. It depends on the characteristics of the source, channel and detectors, and hence is a function of mean photon number $\mu$ and the yield $Y_i$. The overall gain for any protocol is given in terms of yields by: 
\begin{eqnarray}\label{eq:gain}
 Q_{\mu} &=& \sum_{i=0}^{\infty} Y_{i} \frac{\mu^{i}}{i!} e^{-\mu} = \sum_{i=0}^{\infty} Q_{i}
\end{eqnarray}
where,
\begin{eqnarray}\label{eq:gaini}
 Q_i &=& Y_{i} \frac{\mu^{i}}{i!} e^{-\mu} 
\end{eqnarray}
\textbf{Quantum Bit Error Rate (QBER):} Quantum Bit error rate (QBER) $E_{\mu}$ is the rate with which Bob makes a wrong detection even when his basis is compatible with Alice's. The error rate of an \textit{k}-photon state $e_k$ for an ideal protocol is given as,
\begin{equation}\label{eq:11}
    e_{k}=\frac{e_{0} Y_{0}+e_{d} \eta_{k}}{Y_{k}}.
\end{equation}
For any protocol (not necessarily an ideal protocol), QBER is related to mean photon number $\mu$, yields $Y_k$ and error rates $e_k$ as 
\begin{eqnarray}\label{eq:qber}
 E_{\mu} Q_{\mu} &=& \sum_{k}^{\infty} e_{k} Y_{k} \frac{\mu^{k} }{k!} e^{-\mu} 
\end{eqnarray}

\section{Analytic bounds for key-rates} 
For completeness, we discuss the traditional method used to compute analytical bounds on the decoy state protocol. The secure key generation rate for the decoy based protocol is given by the equation
% \begin{equation}\label{eqn: GLLP rate} 
% \begin{split}
%     R &\geq \frac{1}{2} \Big\{ - Q_{\mu} f(E_{\mu})H_\bin(E_{\mu}) +  Q_1 \left( 1 -  H_\bin(e_1) \right) \Big\},
% \end{split}
% \end{equation}
\begin{align}\label{eqn: GLLP rate} 
    R &\geq \frac{1}{2} \Big\{ - Q_{\mu} f(E_{\mu})H_\bin(E_{\mu}) \nonumber \\
    &\qquad + Q_1 \left( 1 -  H_\bin(e_1) \right) \Big\},
\end{align}
The constraints for the problem are identical to those in the optimization problem. The key rate of the protocol is a function of several parameters, including the yields of single photon state $Y_{1}$ (or equivalently, the gain $Q_1$) and the corresponding error rate $e_1$. However, these are not experimentally observable in the QKD protocol. What can be experimentally obtained is the overall gain $Q_{\mu}$ and the Quantum Bit Error Rate (QBER) $E_{\mu}$. The decoy state method achieves secure QKD by providing a method to establish rigorous bounds on the parameters $Q_{1}$ and $e_1$ and hence establish bounds on the secure key rate. Such bounds have been derived by Ma. et. al \cite{Ma2005} and are given in equations \eqref{eq:y0}-\eqref{eq:e1}.  % Using the decoy state method the following bounds have been derived 

\section{Numerically computing key rates}
Here we present a method to compute tight lower bounds for the optimization problem \eqref{eqn: basic optimization problem-1}. 
\subsection{Converting into a finite optimization problem}\label{app: finite optimization problem} 
As the objective function of problem \eqref{eqn: basic optimization problem-1} only consists of $2$ terms with $4$ variables, removing infinite parameters from constraints suffices. Consider the following lower bound on the constraint for $Q_{\nu_d}$ in the optimization problem:
\begin{equation}
\begin{aligned}
    Q_{\nu_d} e^{\nu_d} &= \sum_{k = 0}^{n} Y_{k} \frac{\nu_d^k}{k!} + \sum_{k = n+1}^{\infty} Y_{k} \frac{\nu_d^{k}}{k!} \\
    &\leq \sum_{k = 0}^{n} Y_{k} \frac{\nu_d^k}{k!} + \sum_{k = n+1}^{\infty} \frac{\nu_d^{k}}{k!}
\end{aligned}
\end{equation}

where the above inequality holds because $Y_{k} \in [0, 1]$. Therefore, defining $\Theta_n(\nu_d)$ by the sum

\begin{equation}
    \Theta_n(\nu_d):= \sum_{k = n+1}^{\infty} \frac{\nu_d^{k}}{k!} = e^{\nu_d} - \sum_{k = 0}^{n} \frac{\nu_d^{k}}{k!}
\end{equation} 

allows us to obtain a relaxation corresponding to the constraint $Q_{\mu} e^{\mu} = \sum_{k = 0}^{\infty} Y_{k} \mu^{k}/k!$ in optimization problem \eqref{eqn: basic optimization problem-1}:

\begin{equation}
    Q_{\mu} e^{\mu} - \Theta_{n}(\mu) \leq \sum_{k= 0}^{n} Y_{k} \frac{\mu^{k}}{k!} \leq Q_{\mu } e^{\mu}. \label{eqn: relax1}
\end{equation}

In other words, we have formed a relaxation of the constraint given by an infinite sum of infinitely many variables through a constraint that involves a finite sum of finitely many variables. We use similar arguments to derive the following relaxations of the constraints for $E_{\mu} Q_{\mu}, Q_{\nu_i}, E_{\nu_i} Q_{\nu_i}$ in optimization problem \eqref{eqn: basic optimization problem-1}:

\begin{align}
    &E_{\nu_d} Q_{\nu_d} e^{\nu_d} - \Theta_n(\nu_d) \leq \sum_{k= 0}^{n} e_{k} Y_{k} \frac{\nu_d^{k}}{k!} \leq E_{\nu_d} Q_{\nu_d }  e^{\nu_d} \label{eqn: relax2}
\end{align}

The above relaxations allow us to prove the following lemma:

\begin{lemma}\label{lemm: rn vs r}
Let  $S = \{ \mu  \} \cup \{ \nu_i \}_{i = 1}^{m}$, then consider the optimization problem 
\begin{align}\label{eqn: optimization problem}
r_{n} :=  \inf &  \Big( Y_1 \mu \left( 1 - \Phi\left(2 e_1 -1 \right) \right) \nonumber\\
& +   Y_2 \frac{\mu^2}{2} \left(1 - \Phi\left( (2 e_2 - 1)^2 \right) \right)\Big) \nonumber\\
\textrm{s.t. } &  \forall i \in \{ 0 , \cdots , n \}: \quad Y_{i} , e_{i} \in [0 , 1] \nonumber\\
&  \forall \lambda \in S:   Q_{\lambda} \exp(\lambda) - \Theta_{n}(\lambda) \leq   \sum_{k= 0}^{n} Y_{k} \frac{\mu^{k}}{k!}     \nonumber\\
& \forall \lambda \in S:  \sum_{k= 0}^{n} Y_{k} \frac{\mu^{k}}{k!}  \leq Q_{\lambda} \exp(\lambda) \nonumber\\ 
&   \forall \lambda \in S:  E_{\lambda}Q_{\lambda} \exp(\lambda) - \Theta_{n}(\lambda) \leq   \sum_{k= 0}^{n} e_{k}  Y_{k} \frac{\mu^{k}}{k!}  \nonumber\\ 
&   \forall \lambda \in S:  \sum_{k= 0}^{n} e_{k}  Y_{k} \frac{\mu^{k}}{k!}   \leq E_{\lambda} Q_{\lambda} \exp(\lambda) 
\end{align} 
Then $r_{n} \leq r$. Furthermore, $$ \lim_{n \rightarrow \infty} |r_{n} - r|  = 0 $$
\end{lemma}
\begin{proof} 
Consider the optimization problems for $r_{n}$ and $r$. From the preceding discussion, it is evident that the constraints in problem \eqref{eqn: optimization problem} serve as relaxations for the constraints in problem \eqref{eqn: basic optimization problem-1}. Consequently, the feasible set of optimization problem \eqref{eqn: basic optimization problem-1} is a subset of that of optimization problem \eqref{eqn: optimization problem}. Moreover, the optimization problems designed to compute both $r$ and $r_{n}$ share identical objective functions. Thus, $r_{n} \leq r$ holds for all $n \in \mathbb{R}$.

Additionally, it is straightforward to recognize that when \(n < m\), the constraints of the optimization problem for \(r_{n}\) are relaxations of those for \(r_{m}\). This immediately implies that \(r_{n} \leq r_{m}\). As we continue this process, the feasible set for the optimization problem for \(r_{n}\) approaches that of the optimization problem for \(r\). Consequently, the non-decreasing sequence \(\{ r_{n} \}_{n}\) will converge to \(r\).
\end{proof} 

\subsection{Partitioning the domain}\label{app: partitioning the domain}
In this section, we rigorously define a partition of a set $\mathcal{D}$ and then demonstrate how partitioning the parameter set of an optimization problem can be employed to provide lower bounds for any general optimization problem. We subsequently utilize this technique to lower bound asymptotic key rates in the next section.

Consider an optimization problem of the form: 
\begin{equation}\label{eqn: arbit_optimization_problem}
\begin{aligned}
     k :=  \min&  f(\B{x}, \B{y}) \\ 
     \mathrm{s.t.} & \B{x} \in \M{D}\\ 
      & \B{y} \in \M{Q} \\
      & \forall i \in \{ 1 , \cdots , N\}: g_i(\B{x},\B{y}) \leq 0 \\ 
      & \forall j \in \{ 1 , \cdots , M \}: h_j(\B{x},\B{y}) = 0
\end{aligned}
\end{equation}
where $\M{D} \subset \mathbb{R}^{n}$, $\M{Q} \subset \mathbb{R}^{m}$ and $f,g_1 ,\cdots g_{N}, h_{1} , \cdots h_{M} $ are smooth functions defined on the domain $\M{D} \times \M{Q}$. We seek to compute lower bounds on $k$.

We now formally define a partition of a set $ \M{D} \subset \mathbb{R}^{n}$. 
\begin{definition}[Partition]Let $\B{a} = (a_1 , a_2 , \cdots a_{n}) , \B{b} = (b_1 , b_2, \cdots , b_{n} ) \in \mathbb{R}^{n}$ be two distinct points such that $\forall i: a_{i} \leq b_{i}$. We adopt the notation $\M{C}_{\B{a},\B{b}} := [a_1, b_1] \times [a_2, b_2] \cdots [a_{n} , b_{n}]$, where $\M{C}_{\B{a},\B{b}}$ represents a (hyper) rectangle with $\B{a}$ and $\B{b}$ as the opposite vertices. We refer to a collection of (hyper) rectangles $\M{P} = \{ \M{C}_{\B{a}^{(i)} , \B{b}^{(i)}} \}_{i}$ as a partition (or grid) of the set $\M{D} \subset \mathbb{R}^{n}$ if $\bigcup_{i} \M{C}_{\B{a}_i \B{b}_i} \supseteq \M{D}$.
\end{definition} 

For a partition $\M{P}$, the following result provides a reliable lower bound on $k$.
\begin{lemma}\label{lem: general_technique}
Let $k$ be defined via optimization problem \eqref{eqn: arbit_optimization_problem}. Let $\M{P} = \{ \M{C}_{\B{a}^{(i)} , \B{b}^{(i)} } \}_{i}$ be any partition of the set $\M{D}$. For each $\M{C}_{\B{a}^{(i)}, \B{b}^{(i)}}$, let $\tilde{f}_{i}: \M{C}_{\B{a}^{(i)}, \B{b}^{(i)}} \times \M{Q} \mapsto \mathbb{R}$ be any lower bound of $f$ in the domain $\M{C}_{\B{a}^{(i)} , \B{b}^{(i)}}$. Then corresponding to each (hyper) rectangle $\M{C}_{\B{a}^{(i)}, \B{b}^{(i)}}$, define a sub-optimization problem: 
\begin{equation}\label{eqn: arbit_sub_optimization_problem}
\begin{aligned}
     k(\M{C}_{\B{a}^{(i)},\B{b}^{(i)}}) :=  \min&  \tilde{f}_{i}(\B{x}, \B{y}) \\ 
     \mathrm{s.t.} & \B{x} \in \M{C}_{\B{a}_{i} , \B{b}_{i}}\\ 
     &  \B{y} \in \M{Q} \\
      & \forall i \in \{ 1 , \cdots , N\}: g_i(\B{x},\B{y}) \leq 0 \\ 
      & \forall j \in \{ 1 , \cdots , M \}: h_j(\B{x},\B{y}) = 0
\end{aligned}
\end{equation}
Then the following holds: 
\begin{eqnarray}
    k \geq k_{\M{P}} :=   \underset{i}{\min} \quad\{ k(\M{C}_{\B{a}^{(i)} , \B{b}^{(i)}}) \} 
\end{eqnarray}
Furthermore, if $\underset{{\B{x}  \in \M{D} , \B{y} \in \M{Q}}}{\sup} |f - \tilde{f}| \leq \epsilon $, then $k - k_{\M{P}} \leq \epsilon$.
\end{lemma}
\begin{proof}
Let $\B{x}^{*}, \B{y}^{*}$ be optimal for the problem \eqref{eqn: arbit_optimization_problem}. Then $\B{x}^{*} \in \M{C}_{\B{a}^{(i)}, \B{b}^{(i)}}$ for some $\B{C}_{\B{a}^{(i)}, \B{b}^{(i)}}$. Furthermore, we define $\B{x}_{i}, \B{y}_{i}$ to be the optimal for optimization problem \eqref{eqn: arbit_sub_optimization_problem}. Thus, we have that 
\begin{align}
 k_{\M{P}} \leq k(\M{C}_{\B{a}^{(i)}, \B{b}^{(i)}}) &= \tilde{f}(\B{x}_i , \B{y}_i) \leq \tilde{f}_{i}(\B{x}^{*}, \B{y}^{*} ) \nonumber\\
 &\leq  f(\B{x}^{*} , \B{y}^{*}) = k.  
\end{align} 
Here the first inequality (starting from the left to right) follows from the definition of $k_{\M{P}}$. The first equality and the second inequality  follows from the definition of $\B{x}_{i} , \B{y}_{i}$. The third inequality follows from the fact that $\tilde{f}$ is a lower bound on $f$ in $\M{C}_{\B{a}^{(i)}, \B{b}^{(i)}}$. The final equality follows from the definition of $\B{x}^*, \B{y}^* $. \\ 
To prove the second part, define  
\begin{align}
     k_1(\M{C}_{\B{a}^{(i)},\B{b}^{(i)}}) &:=  \min  f_{i}(\B{x}, \B{y}) \nonumber \\ 
     \mathrm{s.t.} & \B{x} \in \M{C}_{\B{a}_{i} , \B{b}_{i}} \nonumber \\ 
     &  \B{y} \in \M{Q} \nonumber \\
      & \forall i \in \{ 1 , \cdots , N\}: g_i(\B{x},\B{y}) \leq 0 \nonumber \\ 
      & \forall j \in \{ 1 , \cdots , M \}: h_j(\B{x},\B{y}) = 0
\end{align}
Clearly $k = \min_{i} \{k_{1}(\M{C}_{\B{a}^{(i)} , \B{b}^{(i)}}) \}$. However, for every cuboid $\M{C}_{\B{a}^{(j)}, \B{b}^{(j)}}$, we have that $$k_{1}(\M{C}_{\B{a}^{(j)}, \B{b}^{(j)}}) - k(\M{C}_{\B{a}^{(j)}, \B{b}^{(j)}}) \leq f(\B{x}_{j} , \B{y}_{j}) -  \tilde{f}(\B{x}_{j} , \B{y}_{j})  \leq \epsilon. $$ 
here $\B{x}_{j} , \B{y}_{j}$ are such that $k_1(\M{C}_{\B{a}^{(j)}, \B{b}^{(j)}}) = f(\B{x}_{j} , \B{y}_{j})$. \\
This implies that 
\begin{align}
k &= \min_{i} \{k_{1}(\M{C}_{\B{a}^{(i)} , \B{b}^{(i)}}) \} \geq \min_{i} \{k(\M{C}_{\B{a}^{(i)} , \B{b}^{(i)}}) - \epsilon \} \nonumber\\
&\geq \min_{i} \{k(\M{C}_{\B{a}^{(i)} , \B{b}^{(i)}}) \} - \epsilon \geq k_{\M{P}} - \epsilon.
\end{align}
\end{proof}

\subsection{Method for computing key-rates}\label{app: Algo}

As $e_1$ and $e_2$ both vary in the interval $[0,1]$, we construct a partition $\mathcal{P} = \bigcup_{i} \mathcal{C}_{\mathbf{a}^{(i)}, \mathbf{b}^{(i)}}$ of the set $[0,1] \times [0,1]$.

For each cuboid $\mathcal{C}_{\mathbf{a}^{(i)}, \mathbf{b}^{(i)}}$, we define constants $\lambda_{k,i}^{\max}$ as via:
\begin{equation}
\label{eqn: e1_max}
\xi^{\max}_{k,i}= \begin{cases}
    \begin{aligned}
    &\max \{ \Phi((2 a_k^{(i)}  - 1)^{k}) , \\
    &\Phi((2 b_k^{(i)} - 1)^{k}) \} 
    \end{aligned} & \mathrm{if}\quad 0 \notin [2a_k^{(i)} -1 , 2b_k^{(i)} - 1] \\
    \Phi(0)=1 & \mathrm{otherwise}
\end{cases}
\end{equation}
As we shall show below, these constants are defined such that $\xi^{\max}_{k,i}$ tightly upper bound $\Phi((2 e_k - 1)^{k})$ in the rectangle $\mathcal{C}_{\mathbf{a}^{(i)}, \mathbf{b}^{(i)}}$.

\begin{lemma}\label{lem: lowr_bnd}
Let $f( e_1 , e_2 , Y_1 , Y_2) := \sum_{k \in \{ 1 ,2 \} }  Y_{k} \mu^{k}/k! (1 - \Phi((2 e_k - 1)^{k})) $.  Then $f_{i}(Y_1 , Y_2 , e_1 , e_2) := \sum_{k \in \{ 1 ,2 \} }  Y_k \mu^k/k! (1 - \xi^{\max}_{k,i}) $ is a lower bound on $f(e_1 , e_2 , Y_1 , Y_2 )$ in the cuboid $\mathcal{C}_{\B{a}^{(i)} , \B{b}^{(i)} }$.
\end{lemma}

\begin{proof}
As $Y_{k}$ are non-negative, to show that $f_{i}(\mathbf{x} , \mathbf{y}) \leq f(\mathbf{x} , \mathbf{y})$ in $\mathcal{C}_{\mathbf{a}^{(i)} , \mathbf{b}^{(i)} }$, it suffices to show that $\lambda^{\max}_{k, i} $ is an upper-bound of $\Phi((2 e_{k} - 1)^{k})$ (and hence $1- \xi^{\max}_{k,i}$ is a lower-bound of $1 - \Phi((2e_{k} -1)^{k}) $). \\ 
This is done by noticing certain properties of $\Phi(x)$. In particular, it is useful to note that 
\begin{itemize}
    \item  $\Phi'(x) < 0$ if $x \in (0 ,1) $. 
    \item $\Phi'(x) = - \Phi'(x)$.
\end{itemize}
Using these properties we see that  
\begin{eqnarray}
   \Phi'(x^{k}) :=  \frac{\dd \Phi(x^{k})}{\dd x} = k x^{k -1} \Phi'(x^{k})  \nonumber
\end{eqnarray}
Thus, $\Phi'(x^k) $ also has the same two aforementioned properties: 
\begin{itemize}
    \item  $\Phi'(x^{k}) \leq 0$ if $x \in (0 ,1) $. 
    \item  $\Phi'((-x)^{k})  = k(-x)^{k -1} \Phi'((-x)^{k}) = (-1)^{2k - 1} k x^{k -1} \Phi'(x^{k})  =  -\Phi'(x^{k})  $. 
\end{itemize}

To bound $\Phi((2 x - 1)^{k})$, 
Suppose $2 a_k^{(i)} - 1$ and $2 b_k^{(i)} - 1$ both have the same sign (i.e. they are both positive or both negative), then $\Phi((e_{k} - 1)^{k}) \leq \xi^{\max}_{k,i}$ follows from the fact that $\Phi(x^{k})$ is a monotonic function in $x \in (-1 ,0)$ and $x \in (0 ,1)$. If $2 a_k^{(i)} - 1$ and $2 b_k^{(i)} - 1$ have different sign, then it means that $0 \in [2 a_{k}^{(i)} - 1 , 2 b_{k}^{(i)}  - 1 ]$. As $\Phi(0) = 1$, and $\Phi(x^{k}) \leq 1$ implies the bound $\Phi((e_{k} - 1)^{k}) \leq \xi^{\max}_{k,i}$ holds for this case as well. 
\end{proof}
Using the results above, we are now in the position to prove the following result: \\
\begin{theorem}[]\label{thm: main_result_without_uncertainities} 
Let $\M{P} = \{ \M{C}_{ \B{a}^{(i)} , \B{b}^{(i)}} \}_{i}$ be a partition of the set $[0, 1] \times [0 , 1]$ into rectangles defined as in Appendix \ref{app: partitioning the domain}. Let $\xi^{\max}_{1,i}$, and $\xi^{\max}_{2,i}$ be defined according to eqns \eqref{eqn: e1_max} and $\Theta_{n}(\lambda)$ be any upper bound of the sum $\sum_{i= n +1}^{\infty} (\lambda^{i}/i!)$.  Define the optimization problem $r_{n}(\M{C}_{\B{a}^{(i)} , \B{b}^{(i)}})$ be the polynomial optimization problem 

\begin{equation}\label{eqn: app_optimization}
\begin{aligned}
r_n(\M{C}_{\B{a}^{(i)},\B{b}^{(i)}})& :=  
\inf \quad Y_1 \mu \left(1 - \xi^{\max}_{1 , i } \right) + Y_2 \mu^2/2 \left(1 - \xi^{\max}_{2 , i} \right)   \\
\mathrm{s.t.} \quad & e_{1} \in [a_{1} , b_{1}] \\ 
              \quad & e_{2} \in [a_{2} , b_{2}]  \\
              \quad & \forall i \in \{ 0 , \cdots , n \}: \quad Y_{i} \in [0 , 1] \\ 
              \quad & \forall i \in \{ 3 , \cdots , n \}: \quad e_{i} \in [0 , 1] \\ 
              \quad & \forall \lambda \in \{ \mu, \nu_1 , \cdots , \nu_{K} \}: \\
                    &\quad \quad Q_{\lambda} \exp(\lambda) - \Theta_{n}(\lambda) \leq   \sum_{k= 0}^{n} Y_{k} \frac{\mu^{k}}{k!} \\
                    & \quad \quad\leq Q_{\lambda} \exp(\lambda)  \\
              \quad & \forall \lambda \in \{ \mu, \nu_1 , \cdots , \nu_{K} \}:\\
                    & \quad \quad E_{\lambda}Q_{\lambda} \exp(\lambda) - \Theta_{n}(\lambda) \\
                    & \quad \quad \leq   \sum_{k= 0}^{n} e_{k}Y_{k} \frac{\mu^{k}}{k!}   \leq E_{\lambda} Q_{\lambda} \exp(\lambda)\\ 
\end{aligned}
\end{equation} 

then
\begin{equation}\label{eqn: final_result}
 r \geq \underset{i}{\min} \{ r_{n}(\M{C}_{\B{a}^{(i)} , \B{b}^{(i)}}) \}_{i}.
\end{equation} 
Furthermore, for every $\epsilon > 0$, there exists $n$ and a partition $\tilde{\M{P}} = \bigcup_{j}  \tilde{\M{C}}_{\B{a}^{(j)} , \B{b}^{(j)}}$ such that $|r - \underset{j}{ \min} \hspace{0.1 cm} r_{n}(\M{C}_{\B{a}^{(j)} , \B{b}^{(j)}})| \leq \epsilon$.

\end{theorem}
\begin{proof}
The proof begins by first lower-bounding $r$ using $r_{n}$, as established in Lemma \ref{lemm: rn vs r}. Subsequently, we apply Lemma \ref{lem: general_technique} to compute lower bounds on $r_{n}$. Specifically, let $\B{x} = (e_1 , e_2)$ and $\B{y} = (Y_0 , Y_1 , \cdots , Y_{n} , e_0 , e_2 , \cdots , e_{n})$ denote the parameters. Thus, $\M{D} = [0 ,1 ] \times [0,1]$ and $\M{Q} = [0 ,1] \times [0 , 1] \times \cdots \times [0,1]$ must be chosen.

For any partition $\M{P} = \{ \M{C}_{\B{a}^{(i)} , \B{b}^{(i)} } \}_{i}$ of the set $\M{D} =[0 ,1] \times [0, 1]$, by Lemma \ref{lem: lowr_bnd}, the function $f_{i} := \sum_{k \in \{ 1 ,2 \} } Y_{k} \mu^{k}/k! (1 - \xi^{\max}_{k,i})$ serves as a lower bound to the objective function $f =  \sum_{k \in { 1 ,2 } } Y_{k} \mu^{k}/k! (1 - \Phi((2e_{k} - 1)^{k}))$. 
Following Lemma \ref{lem: general_technique} we arrive at the optimization problem \eqref{eqn: app_optimization}. Furthermore, again by Lemma \ref{lem: general_technique} we also conclude that $\underset{i}{\min} \{ r_{n}(\M{C}_{\B{a}^{(i)} , \B{b}^{(i)}}) \}_{i} \leq r_{n}$. Finally, equation \eqref{eqn: final_result} is a consequence of the fact that $r_{n} \leq r$.

Now, let $\M{P}$ be a partition such that each rectangle $\M{C}_{\B{a}^{(i)} , \B{b}^{(i)}}$ is a rectangle of dimensions $\frac{1}{4N_1} \times \frac{1}{4N_2}$ (in other words, we chose a grid with uniform spacing). Note that the grid is chosen such that either $1/2 \notin (2 a_k^{(i)} - 1 , 2 b_k^{(i)} - 1)$. Therefore, $\xi^{\max}_{k , i}$ is either  $2 a_{k}^{(i)}  -1 $ or $2 b_{k}^{(i)} -1$. \\
Now, let $\epsilon > 0$. As $r_{n} \rightarrow r $, chose $n$ such that $|r_{n} - r| \leq \epsilon$. Using monotonicity properties of $\Phi(x) \in [2 a_k^{(i)} -1 , 2b_{k}^{(i)} - 1] $, we obtain $|\Phi(2 x - 1) - \Phi(2 a_{k}^{(i)} - 1) | \leq |\Phi(2 b_{k}^{(i)} - 1)  - \Phi(2 a_{k}^{(i)} - 1)|  $. Similarly $|\Phi(2 x - 1) - \Phi(2 b_{k}^{(i)} - 1) | \leq |\Phi(2 b_{k}^{(i)} - 1)  - \Phi(2 a_{k}^{(i)} - 1)|$. Thus, in the rectangle $\M{C}_{\B{a}^{(i)} , \B{b}^{(i)}}$, we get that 
% \begin{eqnarray}
%     f - \tilde{f}  \leq \sum_{k \in \{ 1 , 2\}}  Y_{k} \frac{\mu^{k}}{k!} |\Phi((2 b_k^{(i)} - 1)^{k}) - \Phi((2 a_{k}^{(i)} - 1)^{k}) |  \leq \chi \sum_{k \in \{ 1 , 2\}} |\Phi((2 b_{k}^{(i)} - 1)^{k}) - \Phi((2 a_{k}^{(i)} - 1)^{k}) |
% \end{eqnarray}
\begin{align*}
    f - \tilde{f}  &\leq \sum_{k \in \{ 1 , 2\}}  Y_{k} \frac{\mu^{k}}{k!} |\Phi((2 b_k^{(i)} - 1)^{k}) - \Phi((2 a_{k}^{(i)} - 1)^{k}) | \\
    & \leq \chi \sum_{k \in \{ 1 , 2\}} |\Phi((2 b_{k}^{(i)} - 1)^{k}) - \Phi((2 a_{k}^{(i)} - 1)^{k}) |
\end{align*}
where $\chi = \max\{ \mu , \mu^2/ 2 \}$. \\ 
Again, using the monotonicity properties of $\Phi(x^k)$, it can be easily verified that $|\Phi((2a_k^{(i)} - 1)) - \Phi((2 b_{k}^{(i)} - 1)) | \leq |\Phi((1 - 1/(2 N_k))^{k})|$. We can choose $N_{k}$ large enough so that $|\Phi((1 - 1/(2 N_k))^{k})| \leq \frac{\epsilon}{2 \chi}$. Thus $f - \tilde{f} \leq \epsilon$ can be chosen for every $\epsilon > 0$ by choosing a suitable $N_1$ and $N_2$. Again from Lemma \ref{lem: general_technique}, we get that $|r_{n} - \underset{i}{\min} \{ r_{n}(\M{C}_{\B{a}^{(i)}, \B{b}^{(i)}}) \} | \leq \epsilon$.
\end{proof}

\subsection{Solving polynomial optimization problems}\label{app: solving the optimization problem} 
The challenge now remains to solve the optimization problem \eqref{eqn: app_optimization}. Notice that, by design, this optimization problem is a polynomial optimization problem. Such problems are not inherently convex optimization problems; however, they can be relaxed to obtain a converging hierarchy of semi-definite programs (SDPs). Interested readers can refer to the excellent slides by Prof. Hamza Fawzi \cite{Hamza} along with Section II of the article \cite{Polynomial_optimization} and find further references for SDP relaxations of polynomial optimization problems. The resulting SDPs can be solved via well-known efficient algorithms.

In this work, we use the NCPOL2SDPA software~\cite{NCPOL2SDPA} to obtain SDP relaxations of polynomial optimization problems. The software directly computes these relaxations given a polynomial optimization problem and then calls an SDP solver directly, outputting the result. The MOSEK software \cite{mosek} was employed to solve the Semi-Definite Programs used for computing the key rates.

\section{Speeding up solving the optimization problem}\label{app: Speeding up the optimization problem}
Note that in the optimization problem in \eqref{eqn: app_optimization}, it is not necessary to partition the entire domain $[0, 1] \times [0, 1]$. This is because not all combinations of $e_1$ and $e_2$ are feasible within the other imposed constraints. Indeed, there is a maximal value of $e_1$ that complies with all the constraints, which was a fundamental element for establishing bounds on the key rate in the seminal paper by Ma et al. \cite{Ma2005}. Consequently, it is sufficient to consider $e_1$ within the reduced range $[0, e_{1}^{\mathrm{up}}]$, where the upper limit $e_{1}^{\mathrm{up}}$ is determined by the following expression:

\begin{equation}
\begin{aligned}
e_{1}^{\mathrm{up}} &
        := \max \quad e_{1} \\
\textrm{s.t.}   \quad & \forall k \in \{ 0 , \cdots , n \}:                   \quad Y_{k} , e_{k} \in [0 , 1] \\
                \quad &  \forall \lambda \in S:  Q_{\lambda} \exp(\lambda) - \Theta_{n}(\lambda) \\
                & \quad \quad \quad \leq   \sum_{k= 0}^{n} Y_{k} \frac{\mu^{k}}{k!}   \leq Q_{\lambda} \exp(\lambda)  \\
              \quad &   \forall \lambda \in S:  E_{\lambda}Q_{\lambda} \exp(\lambda) - \Theta_{n}(\lambda) \\
                & \quad \quad \quad \leq   \sum_{k= 0}^{n} e_{k}Y_{k} \frac{\mu^{k}}{k!}   \leq E_{\lambda} Q_{\lambda} \exp(\lambda)
\end{aligned}
\end{equation}
Similarly, we can compute the upper bound for \( e_{2} \), denoted as \( e_{2}^{\mathrm{up}} \), by formulating and solving the corresponding optimization problem. This procedure allows us to confine the feasible region for \( e_{2} \) within the constraints of the problem:

\begin{equation}
\begin{aligned}
e_{2}^{\mathrm{up}}&:=
            \max \quad e_{2}   \\
\textrm{s.t.}    \quad & \forall k \in \{ 0 , \cdots , n \}: \quad Y_{i} , e_{i} \in [0 , 1] \\ 
                 \quad &  \forall \lambda \in S:  Q_{\lambda} \exp(\lambda) - \Theta_{n}(\lambda) \\
                 & \quad \quad \quad \leq   \sum_{k= 0}^{n} Y_{k} \frac{\mu^{k}}{k!}   \leq Q_{\lambda} \exp(\lambda)  \\
                 \quad &  \forall \lambda \in S:  E_{\lambda}Q_{\lambda} \exp(\lambda) - \Theta_{n}(\lambda) \\
                 & \quad \quad \quad\leq   \sum_{k= 0}^{n} e_{k}Y_{k} \frac{\mu^{k}}{k!}   \leq E_{\lambda} Q_{\lambda} \exp(\lambda)\\ 
              \quad & e_{1} \in [0 , e_{1}^{\mathrm{up}}] 
\end{aligned}
\end{equation}
In practice, we find that \( e_{1}^{\textup{up}} \ll  1 \) and \( e_{2}^{\textup{up}} \ll  1 \) which allows us to save a significant amount of computation time, rendering this method more practical.

\section{Rates in the presence of statistical errors}\label{app: rates in presence of statistical errors}
When estimating key rates, it is crucial to account for the fact that gains and quantum bit error rates (QBERs) are calculated based on finite statistical data. Consequently, there exists a small, albeit non-zero, probability that the experimental estimates of these values may deviate from their true values. To address such uncertainties, a common approach involves introducing a security parameter known as the completeness error.

Formally, the completeness error is defined as the probability that a protocol aborts in an ideal implementation. Ideally, a small completeness error is preferred. The completeness error, in conjunction with the number of rounds in the QKD protocol, is used to establish confidence intervals for experimental statistics. Consequently, the reported rates are determined by the chosen value completeness error. Opting for a smaller completeness error enhances the security of the protocol but results in lower certified rates.

Before proceeding to the discussion on confidence intervals for the statistics on gain and QBER, we recall that expression \eqref{eqn:DW_rate} is only applicable if the coincidences monitored in the protocol align with the expected (theoretically computed) coincidence distribution. In a protocol with finitely many rounds, it becomes necessary to ensure that the
observed (single and multi-photon coincidences) are within a confidence interval from the theoretically expected distribution. This parameter $\epsilon_{\mathrm{stat}}$is pre-declared and ideally should be set as small as possible. The protocol should abort if the observed (single and multi) photon coincidences do not align with the expected values within the confidence
interval $\epsilon_{\mathrm{stat}}$. In this paper, we do not consider the complete analysis of the statistical analysis for coincidence detections and leave it for future work. 

We now return to the discussion on addressing issues of statistical uncertainties in inferring gains and QBERs. For the sake of simplicity in this section, the signal state is labelled as $\mu = \nu_0$. This means that we will not differentiate between the signal state and the decoy state in our analysis for this section, as the approach is similar for both cases. Moreover, the number of decoy states is represented by $K$. Recall that the Alice and Bob generate keys from $\M{A}_{i} \in \{ 0 , 1 \}$ and $\M{B}_i \in \{ 0 , 1 , \perp \}$, where $i \in \mathfrak{h}$ (which is the set of rounds that are not discarded -- i.e. $\mathfrak{h} = \{ i \in \{ 1 ,2 , \cdots , n \}: \M{X}_i = \M{Y}_i \}$). Finally, to keep track of which decoy (or signal ) was sent, we defined the random variable $\M{D}_{i} \in \{ 0, 1, \cdots, K \}$. Furthermore, define the following random variables 
\begin{eqnarray}
    \M{W}_{i}^{\nu_d} = \begin{cases}
        1 &   \mathrm{if }  \M{D}_k = \nu_{d} ,  \M{A}_{i} \ne \M{B}_{i}   \\ 
        0 &   \mathrm{otherwise}
    \end{cases}
\end{eqnarray}
\begin{eqnarray}
    \M{U}_{i}^{\nu_d} = \begin{cases}
        1 &   \mathrm{if }  \M{D}_i = \nu_{d} ,  \M{B}_{i} \ne \perp   \\ 
        0 &   \mathrm{otherwise}
    \end{cases}
\end{eqnarray}
The gains $Q_{\nu}$ are estimated as $Q_{\nu_{d}}^{\mathrm{obs}}$ and $E_{\nu}^{\mathrm{obs}}$ that can be computed as
\begin{eqnarray}
    Q_{\nu_{d}}^{\mathrm{obs}} &:=& \frac{|\{i :  \M{U}_{i}^{\nu_d } = 1\}|}{N_{\nu_d}} \\ 
    E_{\nu_d}^{\mathrm{obs}} &:=&   \frac{|\{  i: \M{W}_{i}^{\nu_d} = 1\}|}{N_{\nu_d}}. 
\end{eqnarray}
Where $|.|$ is used to denote the cardinality of the sets and $N_{\nu_d}$ is the number of rounds for which signal/decoy state $\nu_{d}$ has sent and the round was not discarded (i.e. $\M{X}_i = \M{Y}_i$ was recorded). 
For our protocol, the abort condition is given by: 
\begin{eqnarray}
 Q_{\nu_d}^{\mathrm{obs}} \in [Q_{\nu_d}^{\mathrm{exp}} - \delta_{\nu_d}^Q , Q_{\nu_d}^{\mathrm{exp}} + \delta_{\nu_d}^Q ] \label{eqn: abort 1}\\ 
 E_{\nu_d}^{\mathrm{obs}} \in [E_{\nu_d}^{\mathrm{exp}} - \delta_{\nu_d}^E , E_{\nu_d}^{\mathrm{exp}} + \delta_{\nu_d}^E ] \label{eqn: abort 2}
\end{eqnarray}
where $Q_{\nu_d}^{\mathrm{exp}}$ and $E_{\nu_d}^{\mathrm{exp}}$ are the true values of gains and QBERs, and $\delta_{\nu_d}^{Q}$ and $\delta_{\nu_d}^{E}$ are the values of statistical tolerances that we are willing to admit in the experimentally observed gains and QBERs. The tolerances $\delta_{\nu_{d}}^{Q}$ and $\delta_{\nu_{id}^{Q}}$ should be determined using the completeness error. To do so, we use the Hoeffding's inequality, which we state here for convenience:
\begin{lemma}[Hoeffding's inequality] 
Let $X_i$  be $n$ i.i.d.\ random variables with $a \leq X_i \leq b$, $a, b \in \mathbb{R}$. If $S  = \sum_{i} X_{i}$ and $\mu=\mathbb{E}(S)$. Then for $t > 0$
\begin{eqnarray}
\mathbb{P}(|S - \mu| \geq t) \leq 2 \e^{-\frac{2t^2}{n (b - a)^2 }}\,.
\end{eqnarray}
\end{lemma}
Here $\mathbb{P}$ and $\mathbb{E}$ are the symbols used to denote probability and expected value of events, respectively.
\begin{theorem}
Suppose the honest protocol that behaves in an i.i.d.\ fashion for $n = \sum_{d = 0}^{K} N_{\nu_d}$ rounds. Let $\{ \delta_{\nu_d}^{Q}\}_{d = 0}^{K} $ and $\{\delta_{\nu_d}^{E}\}_{d = 0}^{K}  $ be the accepted tolerances in gains and QBERs that define the abort conditions \eqref{eqn: abort 1} and \eqref{eqn: abort 2}. The probability that an honest protocol aborts is no greater than 
\begin{eqnarray}
 2 \sum_{d = 0}^{K} \left(   e^{-2N_{\nu_d} (\delta_{\nu_{d}}^{Q})^2}  +  e^{-2 N_{\nu_d} (\delta_{\nu_{d}}^{E})^2}  \right) + 2 \epsilon_{\mathrm{stat}}. 
\end{eqnarray}
\end{theorem}
\begin{proof}
The probability that the honest implementation aborts is given by the probability of the event $\mathfrak{A}$ given by  
\begin{eqnarray}  
 \mathfrak{A} &:=& \left( \bigcup_{d} \{ |Q_{\nu_d}^{\mathrm{obs}} - Q_{\nu_d}^{\mathrm{exp}} | \geq \delta_{\nu_d}^{Q} \}\right) \nonumber \\
 &&\cup \left( \bigcup_{d} \{ |E_{\nu_d}^{\mathrm{obs}} - E_{\nu_d}^{\mathrm{exp}} | \geq \delta_{\nu_d}^{E} \}\right) \cup \mathfrak{A}_1. 
\end{eqnarray}
Where $Q_{\nu_d}^{\mathrm{exp}}$ and $E_{\nu_d}^{\mathrm{exp}}$ are the true values of Gain and QBER and $\mathfrak{A}_1$ is the event when observed values for coincidences are outside the confidence interval from ideal value, despite the true values being within the confidence interval. This can happen almost with the probability $2 \epsilon_{\mathrm{stat}}$. 

We can get a bound on $\mathbb{P}(\mathfrak{A})$ using the union bound
\begin{align*}
    \mathbb{P}(\mathfrak{A}) &\leq \sum_{d} \Bigl( \mathbb{P}(\{ |Q_{\nu_d}^{\mathrm{obs}} - Q_{\nu_d}^{\mathrm{exp}} | \geq \delta_{\nu_{d}}^{Q} \}) \nonumber \\
    &\quad+  \mathbb{P}(\{ |E_{\nu_d}^{\mathrm{obs}} - E_{\nu_d}^{\mathrm{exp}} | \geq \delta_{\nu_{d}}^{E} \})\Bigr) + \mathbb{P}(\mathbb{A}_1) .
\end{align*}
We can bound the above by using Hoeffding's inequality by taking $S = \sum_{i} \M{W}_{i}$ or $S = \sum_{i} \M{U}_{i}$:
\begin{eqnarray}
    \mathbb{P}(\{ |Q_{\nu_d}^{\mathrm{obs}} - Q_{\nu_d}^{\mathrm{exp}} | \geq \delta_{\nu_{d}}^{Q} \}) \leq 2 e^{-2n (\delta_{\nu_{d}}^{Q})^2} \\ 
     \mathbb{P}(\{ |E_{\nu_i}^{\mathrm{obs}} - E_{\nu_d}^{\mathrm{exp}} | \geq \delta_{\nu_{d}}^{E} \}) \leq 2 e^{-2 n (\delta_{\nu_{d}}^{E})^2 }.
\end{eqnarray}
Therefore, the following holds: 
\begin{eqnarray}
  \mathbb{P}(\mathfrak{A}) &\leq& 2 \sum_{s = 0}^{K} \left(   e^{-2N_{\nu_d} (\delta_{\nu_{d}}^{Q})^2}  +  e^{-2 N_{\nu_i} (\delta_{\nu_{d}}^{E})^2}  \right) \nonumber \\
  &&+ 2\epsilon_{\mathrm{stat}}.   
\end{eqnarray}
\end{proof}
Using the result above, we can get a relationship between the completeness error $\epsilon_{C}$ and $\delta_{\nu_d}^{Q}$ or $\delta_{\nu_d}^{E}$ by the setting 
\begin{eqnarray}
 \forall i: \quad    e^{-2 N_{\nu_d} (\delta_{\nu_d}^{Q})^2} =  e^{-2N_{\nu_d} (\delta_{\nu_d}^{E})^2} = \frac{\epsilon_{C} - 2 \epsilon_{\mathrm{stat}}}{2 (K+1)}. 
\end{eqnarray}
Therefore, we can set  
\begin{eqnarray}\label{eqn: delta}
    \delta_{\nu_d}^{Q} = \delta_{\nu_d}^{E} =  \sqrt{\frac{-1}{2 N_{\nu_d}}\ln\left(\frac{\epsilon_{C} - 2 \epsilon_{\mathrm{stat}}}{2 (K+1)} \right)}.
\end{eqnarray}
Using this, we revise the constraints \ref{eqn: relax1} and \ref{eqn: relax2} to the following constraints
% \begin{eqnarray}
% (Q_{\nu_d} - \delta_{\nu_d}^{Q})e^{\nu_d} - \Theta_n(\nu_d) &\leq&   \sum_{k= 0}^{n} Y_{k} \frac{\nu_d^{k}}{k!}   \leq (Q_{\nu_d} + \delta_{\nu_d}^{Q}) e^{\nu_d} \label{eqn: EAT1}\\ 
% (E_{\nu_d} - \delta_{\nu_d}^{E}) (Q_{\nu_d} - \delta_{\nu_d}^{Q}) e^{\nu_d} - \Theta_n(\nu_d) &\leq&   \sum_{k= 0}^{n} e_{k} Y_{k} \frac{\nu_d^{k}}{k!}   \leq (E_{\nu_d}+\delta_{\nu_{i}}^{E})(Q_{\nu_d }+ \delta_{\nu_{d}^{Q}})  e^{\nu_d}. \label{eqn: EAT2} 
% \end{eqnarray}
\begin{widetext}
\begin{align}
(Q_{\nu_d} - \delta_{\nu_d}^{Q})e^{\nu_d} - \Theta_n(\nu_d) &\leq \sum_{k=0}^{n} Y_{k} \frac{\nu_d^{k}}{k!} \leq (Q_{\nu_d} + \delta_{\nu_d}^{Q}) e^{\nu_d} \label{eqn: EAT1} \\
(E_{\nu_d} - \delta_{\nu_d}^{E}) (Q_{\nu_d} - \delta_{\nu_d}^{Q}) e^{\nu_d} - \Theta_n(\nu_d) &\leq \sum_{k=0}^{n} e_{k} Y_{k} \frac{\nu_d^{k}}{k!} \leq (E_{\nu_d}+\delta_{\nu_{i}}^{E})(Q_{\nu_d }+ \delta_{\nu_{d}^{Q}}) e^{\nu_d}. \label{eqn: EAT2}
\end{align}
\end{widetext}
The constraints for Theorem \ref{thm: main_result_without_uncertainities} should be modified accordingly to compute key-rate in the presence of experimental uncertainties. Note that by doing so, we receive a lower key-rate, as \eqref{eqn: EAT1} and \eqref{eqn: EAT2} are relaxations of \ref{eqn: relax1} and \ref{eqn: relax2} respectively.

\bibliographystyle{unsrt}
\bibliography{main}

\end{document}